\documentclass[journal]{IEEEtran}
\usepackage{gensymb}
\usepackage{amsmath}
\usepackage{graphicx}
\usepackage{subcaption}
\usepackage{subfig}
\usepackage{multicol}
\usepackage{booktabs}
\usepackage{tabularx}
\usepackage{url}
\usepackage{amssymb}  % For the checkmark symbol (\checkmark)
\usepackage{multirow} % For multi-row cells in the table
\usepackage{array}    % For better control over column formatting
\usepackage{algorithm}
\usepackage{algorithmic}
\usepackage[numbers]{natbib}

\counterwithin*{equation}{subsubsection}
\begin{document}
\title{Toward Flare-Free Images: A Survey}
\author{Yousef~Kotp, Marwan~Torki\\ \textit{Computer and Systems Engineering Department, Alexandria University, Egypt.}}

\maketitle

\begin{abstract}
Lens flare is a common image artifact that can significantly degrade image quality and affect the performance of computer vision systems due to a strong light source pointing at the camera. This survey provides a comprehensive overview of the multifaceted domain of lens flare, encompassing its underlying physics, influencing factors, types, and characteristics. It delves into the complex optics of flare formation, arising from factors like internal reflection, scattering, diffraction, and dispersion within the camera lens system. The diverse categories of flare are explored, including scattering, reflective, glare, orb, and starburst types. Key properties such as shape, color, and localization are analyzed. The numerous factors impacting flare appearance are discussed, spanning light source attributes, lens features, camera settings, and scene content. The survey extensively covers the wide range of methods proposed for flare removal, including hardware optimization strategies, classical image processing techniques, and learning-based methods using deep learning. It not only describes pioneering flare datasets created for training and evaluation purposes but also how they were created. Commonly employed performance metrics such as PSNR, SSIM, and LPIPS are explored. Challenges posed by flare's complex and data-dependent characteristics are highlighted. The survey provides insights into best practices, limitations, and promising future directions for flare removal research. Reviewing the state-of-the-art enables an in-depth understanding of the inherent complexities of the flare phenomenon and the capabilities of existing solutions. This can inform and inspire new innovations for handling lens flare artifacts and improving visual quality across various applications. 
\end{abstract}

\section{Introduction}
\label{sec:introduction}
\IEEEPARstart{L}{ENS} flare \cite{flare1,fourier,seibert1985removal}, also known as image flare is a common problem that occurs when a camera lens is pointed at a strong light source. It can manifest as ghosting, blooming, or other artifacts that can degrade the image quality. It poses a significant challenge in both photography and computer vision domains. It occurs when intense light sources, such as the sun or powerful artificial lights, directly enter a camera's lens. Instead of faithfully capturing the intended scene, the camera ends up with unwanted artifacts in the form of blurry and multicolored spots or streaks scattered across the image. These artifacts not only obscure the image's content but also degrade its overall visual quality. The presence of lens flare can be particularly problematic for computer vision systems \cite{industrial_survey, computer_vision_survey_systems}, as these unexpected aberrations can fool algorithms designed for image analysis and interpretation. Consequently, addressing lens flare has become a crucial preprocessing step in computer vision applications, paving the way for more accurate and reliable results in tasks such as stereo matching \cite{stereo_matching1,stereo_matching2}, optical flow estimation \cite{optical_flow, optical_flow_intro, optical_flow_deep_learning}, semantic segmentation \cite{semantic_segmentation, semantic_segmentation_2, semantic_segmentation_3, semantic_segmentation_4}, and object detection \cite{object_detection, object_detection_2, object_detection_3, object_detection_4}. The survey's contributions can be summarized as follows:
\begin{figure}[t]
    \begin{minipage}[b]{.48\linewidth}
      \centering
      \centerline{\includegraphics[width=4.5cm]{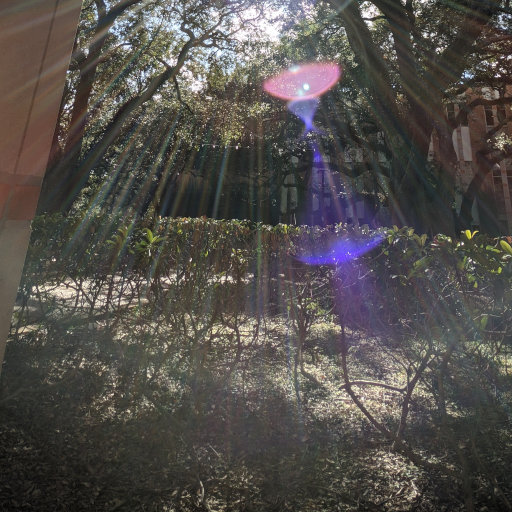}}
      \vspace{0.1cm}
      \centerline{\includegraphics[width=4.5cm]{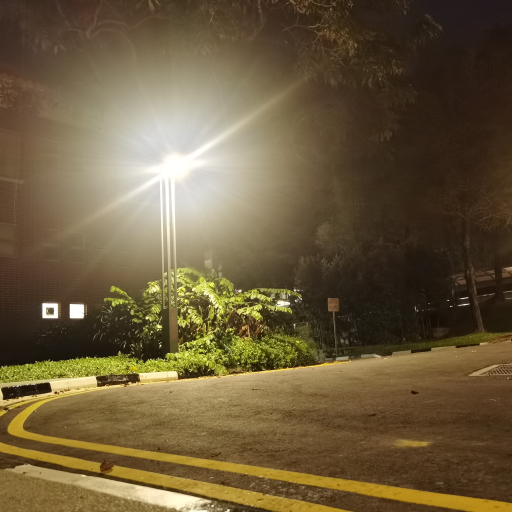}}
      \vspace{0.1cm}
      \centerline{\includegraphics[width=4.5cm]{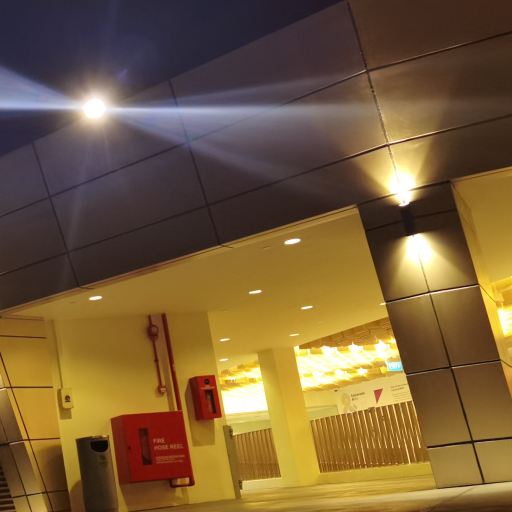}}
    \end{minipage}
\hfill
    \begin{minipage}[b]{0.48\linewidth}
      \centering
      \centerline{\includegraphics[width=4.5cm]{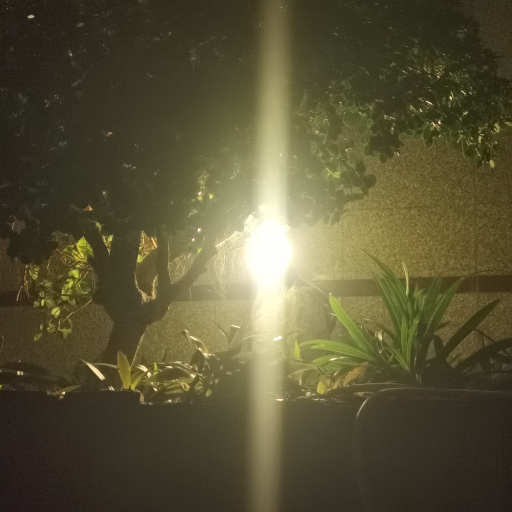}}
      \vspace{0.1cm}
      \centerline{\includegraphics[width=4.5cm]{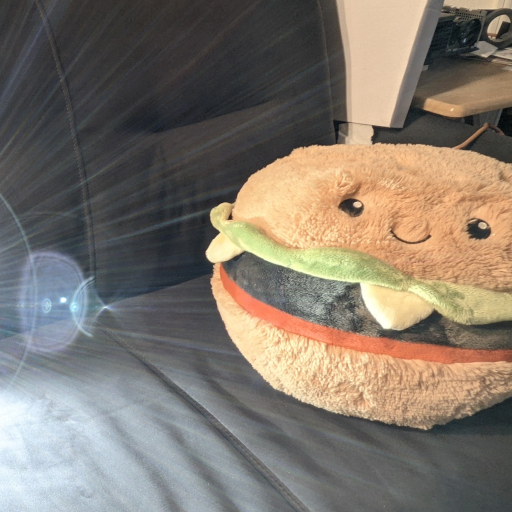}}
      \vspace{0.1cm}
      \centerline{\includegraphics[width=4.5cm]{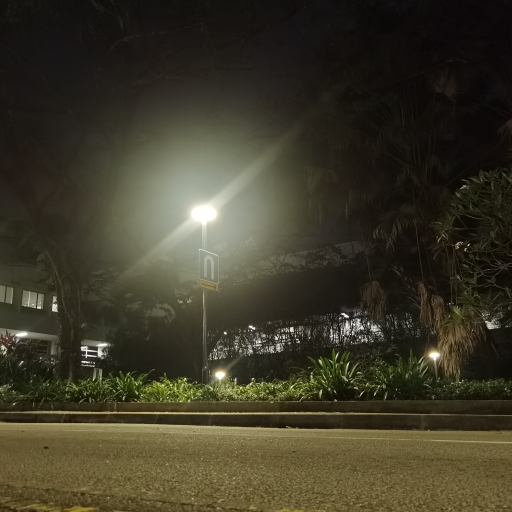}}
    \end{minipage}
\caption{Examples for flare-corrupted images.}
\label{fig:figure1}
\end{figure}
\begin{figure*}[ht]
 \begin{subfigure}[b]{0.195\textwidth}
  \includegraphics[width=\linewidth, height=4cm]{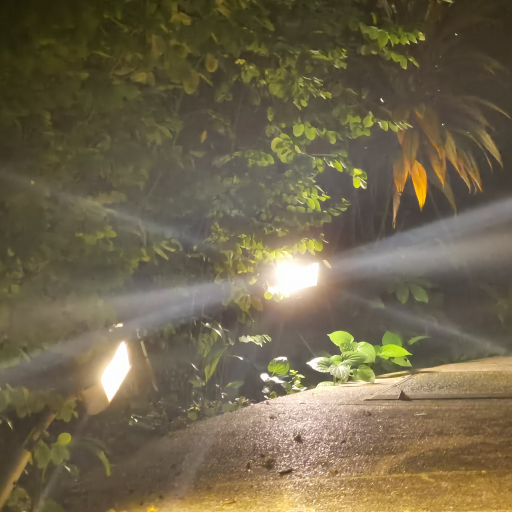}
  \\
  \vspace{-0.15cm}
  \includegraphics[width=\linewidth, height=4cm]{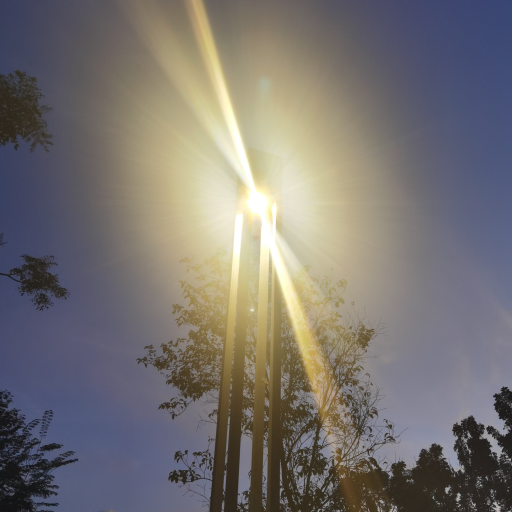}
  \caption{Scattering Flare}
 \end{subfigure}
 \begin{subfigure}[b]{0.195\textwidth}
  \includegraphics[width=\linewidth, height=4cm]{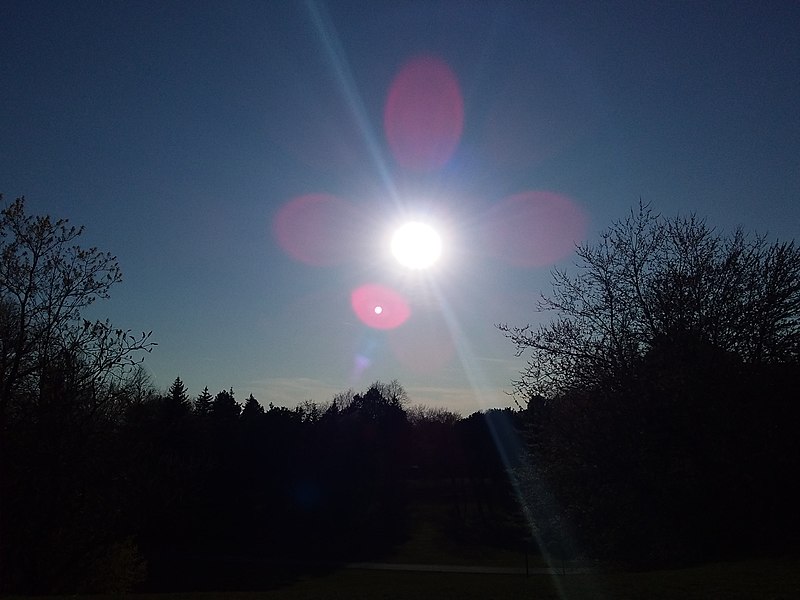}
  \\
  \vspace{-0.15cm}
  \includegraphics[width=\linewidth, height=4cm]{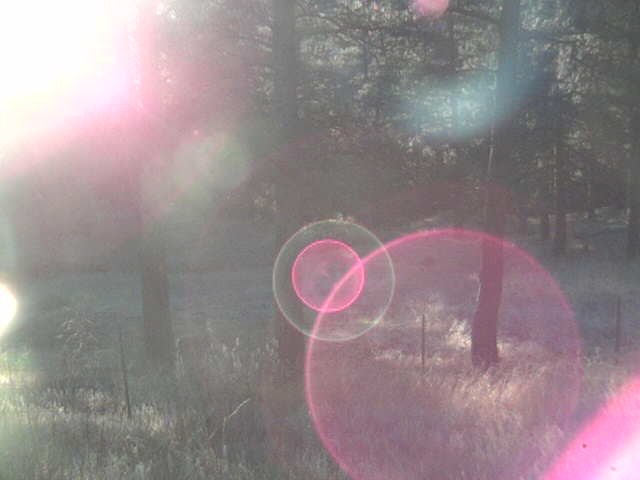}
  \caption{Reflective Flare}
 \end{subfigure}
 \begin{subfigure}[b]{0.195\textwidth}
  \includegraphics[width=\linewidth, height=4cm]{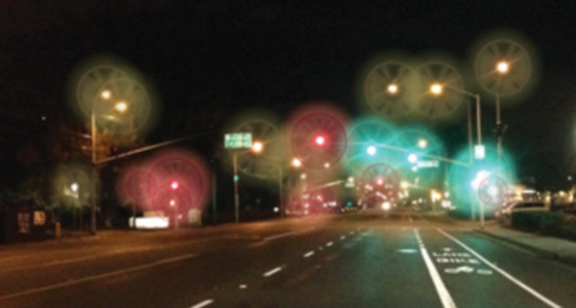}
  \\
  \vspace{-0.15cm}
  \includegraphics[width=\linewidth, height=4cm]{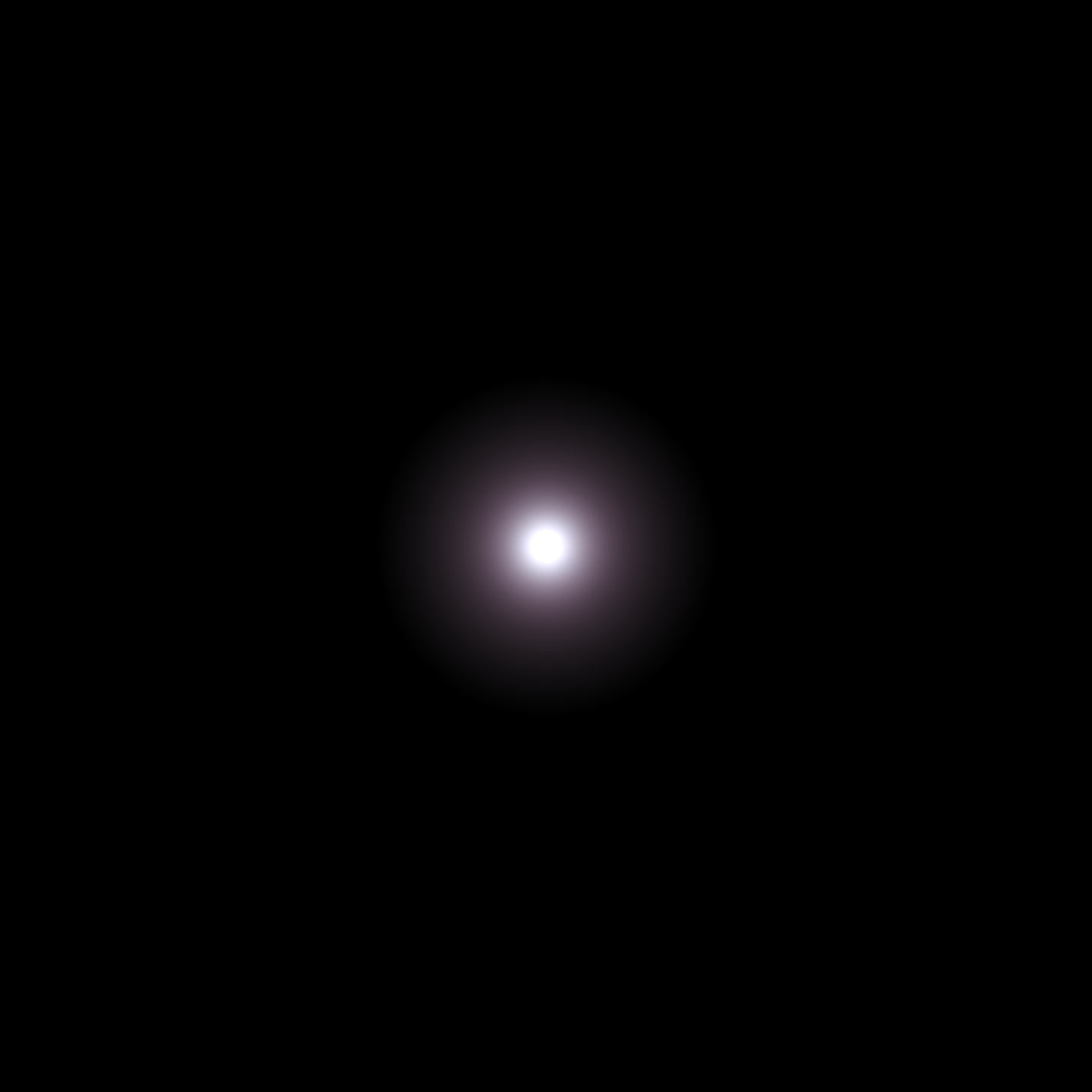}
  \caption{Veiling Glare}
 \end{subfigure}
 \begin{subfigure}[b]{0.195\textwidth}
  \includegraphics[width=\linewidth, height=4cm]{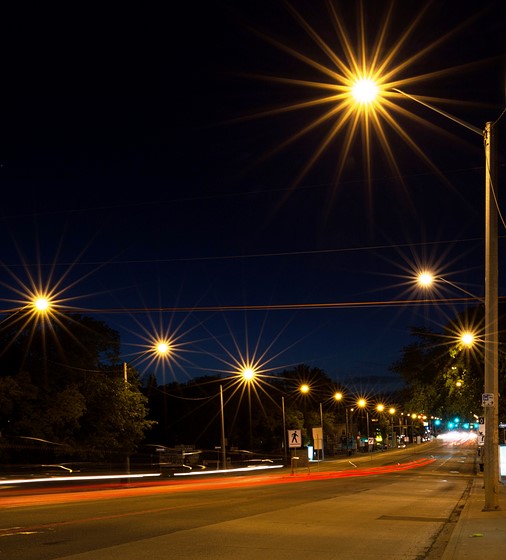}
  \\
  \vspace{-0.15cm}
  \includegraphics[width=\linewidth, height=4cm]{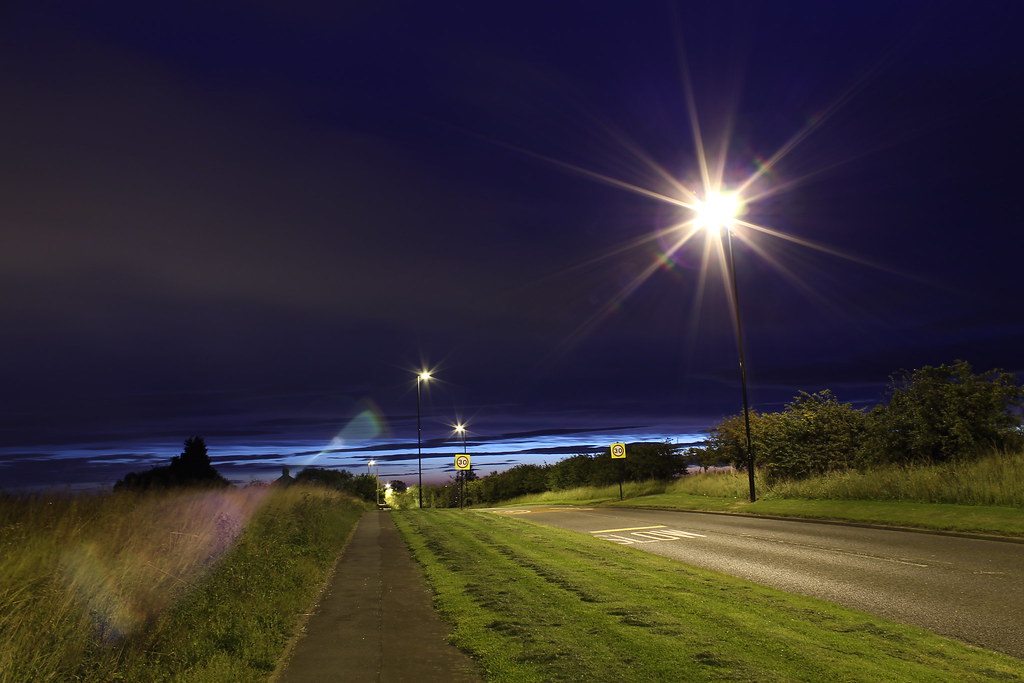}
  \caption{Lens Starburst}
 \end{subfigure}
 \begin{subfigure}[b]{0.195\textwidth}
  \includegraphics[width=\linewidth, height=4cm]{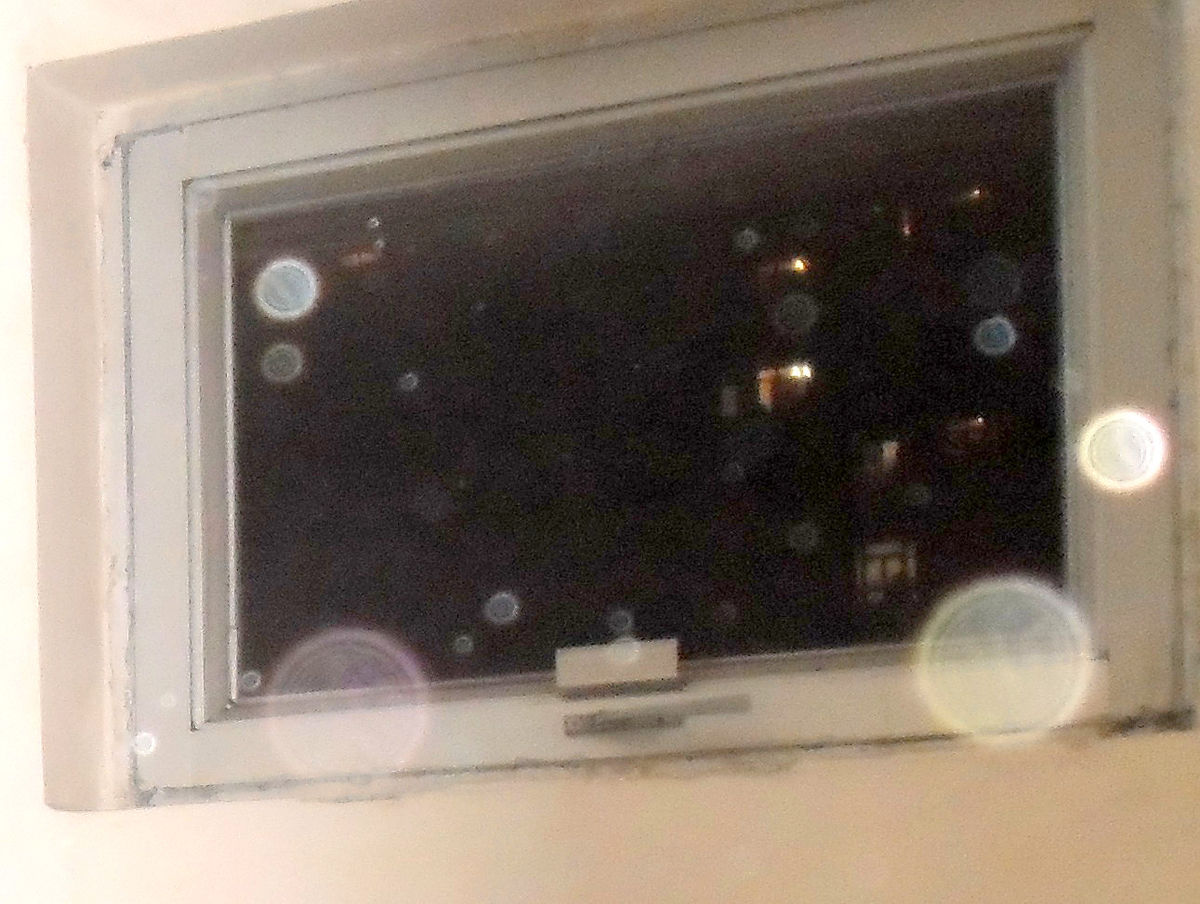}
  \\
  \vspace{-0.15cm}
  \includegraphics[width=\linewidth, height=4cm]{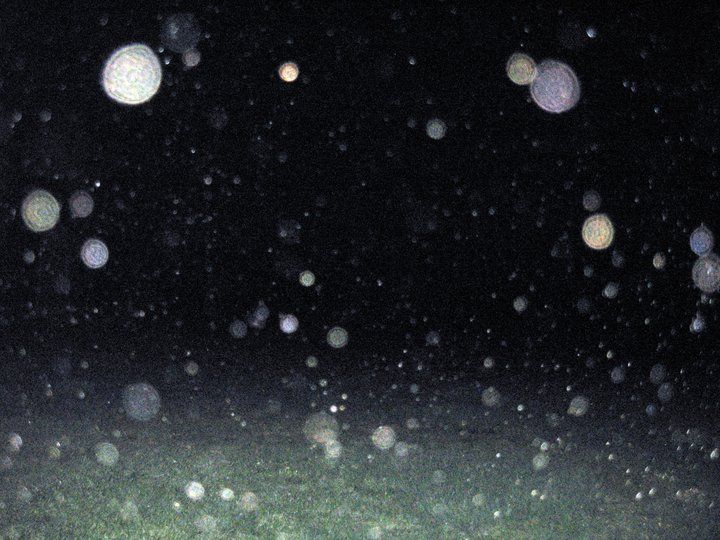}
  \caption{Lens Orbs}
 \end{subfigure}
 \caption{Example for different types of flares. In (a), we can see that scattering flare manifests in the form of a line or radial pattern emerging from the light source, we can also deduce that the effect of scattering flare can be magnified during nighttime. (b) shows the effect of reflective flare in the image which is usually an aperture-shaped polygon, its color can also vary depending on the type and color of lens coating which was used. From (c), we can notice what glare represents, which is induced by excessive and uncontrolled brightness or intensity of light that can cause discomfort and reduced visibility. It occurs when there is a significant contrast between a bright light source and the surrounding environment, often resulting in a dazzling or blinding effect. It is worth noting that scattering flare usually includes glare around the light source. In (d), we can clearly notice the star-shaped pattern around the light source, hence the name starburst. (e) represents lens orbs, which manifest in the form of ghost-like shapes within the image. These ghostly manifestations often depend on particulate matter within the lens optical system. They can introduce dreamy quality to the photograph depending on the scene context.}
 \label{fig:fig2}
\end{figure*}
\begin{itemize}
    \item We provide a comprehensive analysis of the physics, optics, and factors behind lens flare formation. In addition to this, we categorize different flare types according to their patterns.
    \item We discuss different challenges posed by the complex, data-dependent nature of lens flares that make them difficult to model and remove effectively.
    \item We extensively review different methods for lens flare removal, covering hardware-based strategies, classical image processing techniques, and learning-based methods using deep neural networks.
    \item We review pioneering flare datasets and how they were created for training and evaluation purposes.
    \item We provide a comparative summary of the performance of the reviewed flare removal methods on popular benchmarks.
    \item We highlight promising future research directions such as flare removal in videos, real-time processing, improved datasets, modular approaches, daytime-to-nighttime translation, and multi-stage methods.
\end{itemize}

The paper is structured as follows: Section \ref{sec:physics of lens flare} delves into the physics of lens flare, covering its causes, types, and influencing factors. Section \ref{sec:hardware-based} explores hardware-based approaches for flare mitigation, while Section \ref{sec:computational-based} investigates computational methods for flare removal. In section \ref{sec:learning-based}, state-of-the-art learning-based approaches using deep learning are discussed. Section \ref{sec:flare datasets} details the creation of flare removal datasets, and section \ref{sec:performance review} provides an extensive quantitative analysis of the discussed methods. Finally, in section \ref{sec:challenges and opporutnities}, current limitations and future research directions in the field of flare removal are addressed.

\begin{figure}[h]
    \centering
    \includegraphics[width=\linewidth]{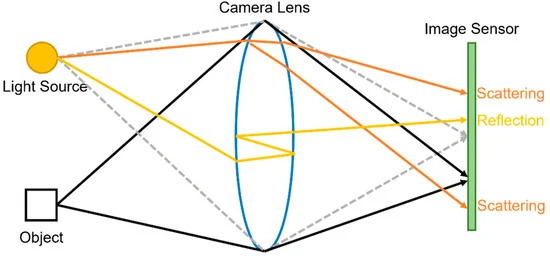}
    \caption{Causes of two main types of flare: scattering and reflective. Scattering flare happens due to the dispersion of light within the camera's optical system, often resulting from refractions of incoming light to unintended positions on the camera sensor. On the other hand, reflective flare occurs when intense light sources, such as the sun or bright artificial lights, directly illuminate the camera lens, causing reflections at various lens elements. It can also happen due to scratching the lens surface. Those internal reflections result in unwanted polygon-shaped artifacts. It is worth noting that this figure contains only one lens for simplicity, in a typical camera system, multiple lenses are commonly employed. From \cite{physicsflare}.}
    \label{fig:physics_of_lens_flare}
\end{figure}

\section{Physics of Lens Flare}
\label{sec:physics of lens flare}
When you first capture an image \cite{digital_image_processing, physics_of_camera}, light from the scene travels through the camera's lens system and eventually reaches the camera's image sensor. The primary function of the camera lens is to bend and focus incoming light onto the image sensor to create a sharp and well-defined image. However, lens flare occurs when certain conditions are met during this light propagation. In reality, actual camera lenses tend to disperse and bounce light in unintended directions, giving rise to unwanted flare artifacts. You can see an example of these flare artifacts in Figure \ref{fig:figure1}. The portions of light that get scattered represent only a small fraction of the total incoming light rays. Consequently, in most photos, this flare is so minor that it goes unnoticed. However, things change when a powerful light source, like the sun, is significantly brighter than the rest of the scene being captured. In such cases, even the small portion of light that gets scattered by this intense source can create visible artifacts on other parts of the image. The particular patterns created by the scattering of light due to dust and scratches result in distinct visual characteristics. The detailed factors contributing to the flare are discussed in section \ref{sec:factors affecting the flare}.

One of the other fundamental physical principles behind lens flare is the phenomenon of internal reflection within the camera lens. A lens consists of multiple optical elements, including curved glass or plastic surfaces. Some of the light rays are inevitably scattered and reflected within the lens elements. This happens due to many factors as the incident angle at which the intense light enters the camera lens is a crucial factor. Light rays can bounce back and forth between the lens elements, creating a series of reflections and scattering. In addition to this, the design of the lens including the number and arrangement of lens elements, can influence the likelihood of lens flare. Complex lens designs with multiple elements may be more prone to internal reflections.

The shape and position of the aperture blades within the lens can also affect the appearance of lens flare. The aperture creates multiple points of reflection for incoming light, potentially resulting in polygonal or starburst-like flare patterns. External factors such as humidity and airborne particles in the atmosphere can scatter and diffract light too, adding to the complexity of the flare pattern. These factors can contribute to the colorful and unpredictable nature of lens flare. The lens flare artifacts that we observe in images are a consequence of these internal reflections and scattering processes. The reflections and scattering create bright spots, streaks, polygons, or other patterns in the image, which can obscure the original scene's details and reduce image contrast.

Figure \ref{fig:physics_of_lens_flare} demonstrates a perfect example of how flare can occur in photography \cite{physicsflare}. Flare arises from two primary causes as discussed previously. One is due to light dispersion within the camera's optical system, typically resulting from the refraction of incoming light to unintended positions on the camera sensor (scattering flare). This is exemplified by the orange light ray in the figure. The other cause is when strong light sources, such as the sun or bright artificial lights, directly illuminate the camera lens, leading to reflections at various lens elements (reflective flare), as shown by the yellow light ray. Those internal reflections produce undesirable polygon-shaped artifacts.

\subsection{Flare Types}
\label{sec:flare types}
Lens flares manifest in images in a variety of shapes and patterns based on the conditions of light interaction within the lens system as mentioned in the previous subsection. They can be broadly categorized into the following types as shown by Figure \ref{fig:fig2}:

\textbf{Scattering Flare:} \cite{scattering1,scattering2,scattering3,scattering4} These appear as radiating streaks spreading outward from bright light sources. They are caused by light scattering off defects, scratches, or dirt particles on the lens surface. The streak patterns tend to remain fixed relative to the light source when the camera moves.

\textbf{Reflective Flare} \cite{reflective1,reflective2,reflective3,reflective4} These are caused by internal reflections between the lens elements. They often appear as a string of polygonal shapes such as circles, ovals, or hexagons. In contrast to scattering flares, these shapes tend to move in the direction opposite to the direction of light source movement.

\textbf{Veiling Glare:} \cite{assessment_of_glare, scattering3} This causes an overall foggy or hazy region around bright light sources. It is caused by the forward scattering of light within the lens system and leads to reduced local image contrast. Most of the published research considers glare a part of the scattering flare.

\textbf{Lens Starburst:} \cite{charach_lens_starburst} These manifest as bright streaks radiating from light sources forming a star-like pattern. They are caused by diffraction and scattering from the aperture blades. Since they exhibit nearly the same streak pattern as scattering flare, researchers consider lens starburst a part of the scattering flare.

\textbf{Lens Orbs:} These appear as out-of-focus blobs that do not move with camera motion. They are mainly caused by particulate matter within the lens system.

Due to the fixed location of lens orbs, they can be removed easily, that's why nearly all of the papers that are published nowadays address only two types of flare: scattering and reflective. Researchers also implicitly consider both glare and lens starbursts as part of the scattering flare, as mentioned previously.

\subsection{Factors Affecting The Flare}
\label{sec:factors affecting the flare}
Several interconnected factors can contribute to the occurrence and characteristics of lens flare. Some of these are related to the camera capturing the scene, and others are related to the environment regardless of the camera. We can mainly narrow the factors affecting the flare to the following:

\textbf{Light Source Properties:} The intensity, color, shape, and positioning of the light source impacts flare patterns. Brighter light causes more prominent flares. Colored lights lead to tinted flare streaks. The shape and orientation of extended sources like neon signs determine flare streak geometry.

\textbf{Lens Properties:} The number of lens elements, their curvature, spacing, and anti-reflective treatments influence internal reflection paths that cause flares. These treatments are explained in Section \ref{sec:hardware-based}. Also, the lens aperture shape drives patterns like polygonal shapes. Lens defects, scratches, and dust lead to scattering artifacts.

\textbf{Camera Settings:} Exposure time, aperture size, and ISO sensitivity affect flare prominence. Longer exposures make flares more noticeable. Higher f-numbers increase diffraction causing starburst patterns.

\textbf{Scene Content:} Flare visibility depends on the scene. Flares are more obvious against dark regions as in nighttime photography compared to textured backgrounds. The presence of other bright regions can mask flare patterns.

\textbf{Shooting Conditions:} Camera angle relative to the light source and distance impacts flare shapes and locations. Environmental factors like humidity affect dust buildup on lenses.

\section{Hardware-Based Approaches}
\label{sec:hardware-based}
Mitigating flare through hardware optimization aims to prevent flare by reducing the need for post-processing. It involves the use of physical components and mechanisms within the camera and lens system to reduce or eliminate the unwanted effects of lens flare. The following are some hardware-based methods to reduce the flare.
\subsection{Anti Reflective Coating}
\begin{figure}[t]
    \centering
    \includegraphics[width=\linewidth]{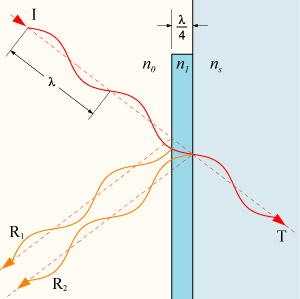}
    \caption{Light passing through both mediums ends up being reflected twice. First when it hits the surface of the AR coat, and second when it hits the surface of the object. Thus we have two distinct rays R1 and R2 traveling in the same direction away from the object. Interference from two similar waves traveling in the same direction is a regular wave – that depends solely on the phase difference of the two waves. The phase difference is caused by the fact that ray R1 travels a distance shorter by 2w than ray R2 where w is the width of the coating layer. Different phase differences produce different merged waves. When the two waves are separated by half the wavelength. The resultant wave has zero amplitude. At every point, the two waves cancel each other. Therefore, if we give the lens a coat of transparent paint of a width of a quarter of the wavelength of the light, then R2 will be out of phase by half a wavelength and therefore produce destructive interference with R1. From \cite{destructive_wave}.}
    \label{fig:physics of ar coating}
\end{figure}
\begin{figure}[t]
    \centering
    \includegraphics[width=\linewidth, height = 3cm]{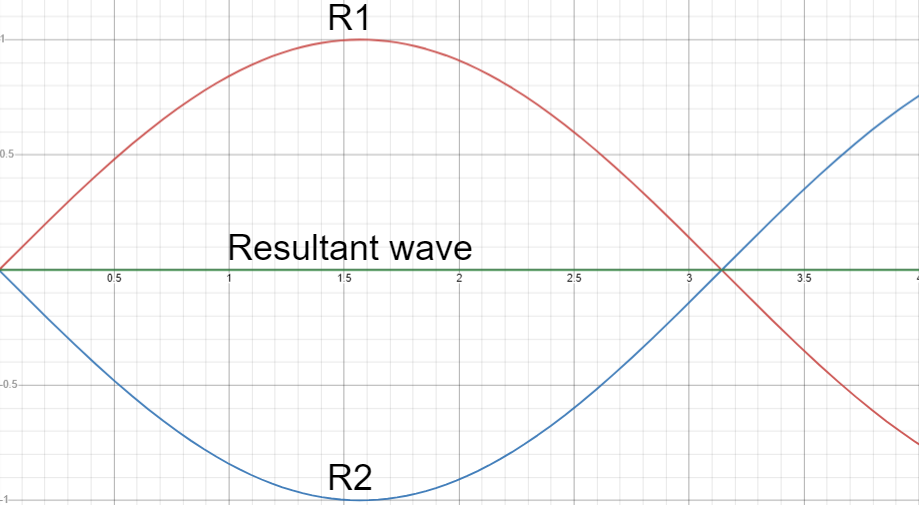}
    \caption{Destructive interference is produced when 2 waves are out of phase by half a wavelength.}
    \label{fig:Destructive Interference}
\end{figure}

Anti-reflective coating \cite{reduction_of_reflections_for_camera_lenses}, often referred to as AR coating, is a thin layer of special material applied to the surfaces of optical components like lenses and camera filters to reduce unwanted reflections and glare. The purpose of AR coating is to enhance the transmission of light through the lens or filter while minimizing the amount of light that is bounced back. With nothing added to the surface of a glass lens, about 4\% of the light hitting it gets reflected back and is lost due to differences in the refractive indices between the air and the material of the surface. This reflected light can cause flare and reduce the overall image quality. A camera lens with only a single element has two air-glass surfaces, so the reflection losses double. Modern lenses often have up to 20 elements in them, so uncoated surfaces would result in huge light losses.

The mechanism behind AR coatings involves the principle of destructive interference \cite{destructive_wave} as shown in Figure \ref{fig:Destructive Interference}. By applying a carefully calculated and precisely engineered thin layer of material onto the lens surface, the coating thickness is adjusted to create a phase difference in the reflected light waves. When the reflected light waves interact with each other, they cancel out and reduce the overall reflection. The physics behind AR coating is explained clearly in Figure \ref{fig:physics of ar coating}.

This results in improved image contrast allowing for better differentiation between light and dark areas. In addition to this, it reduces glare and reflections which means that more of the intended light reaches the imaging sensor, resulting in sharper and clearer images. In low-light situations, AR-coated lenses can capture more of the available light, allowing for better performance without introducing excessive glare. The wavelength in visible light is around 500nm, and lens coatings are typically 100nm to 250nm thin layers. To put this into perspective, an average human hair is about a thousand times thicker. However, the thickness of this coating can only be optimized for particular wavelengths and angles of incidence and therefore cannot be perfect. Additionally, adding an AR coating to all optical surfaces is expensive, and may interfere with other coatings as anti-scratch and anti-fingerprint.

\subsection{Lens Hoods}
\begin{figure}[t]
    \centering
    \includegraphics[width=\linewidth]{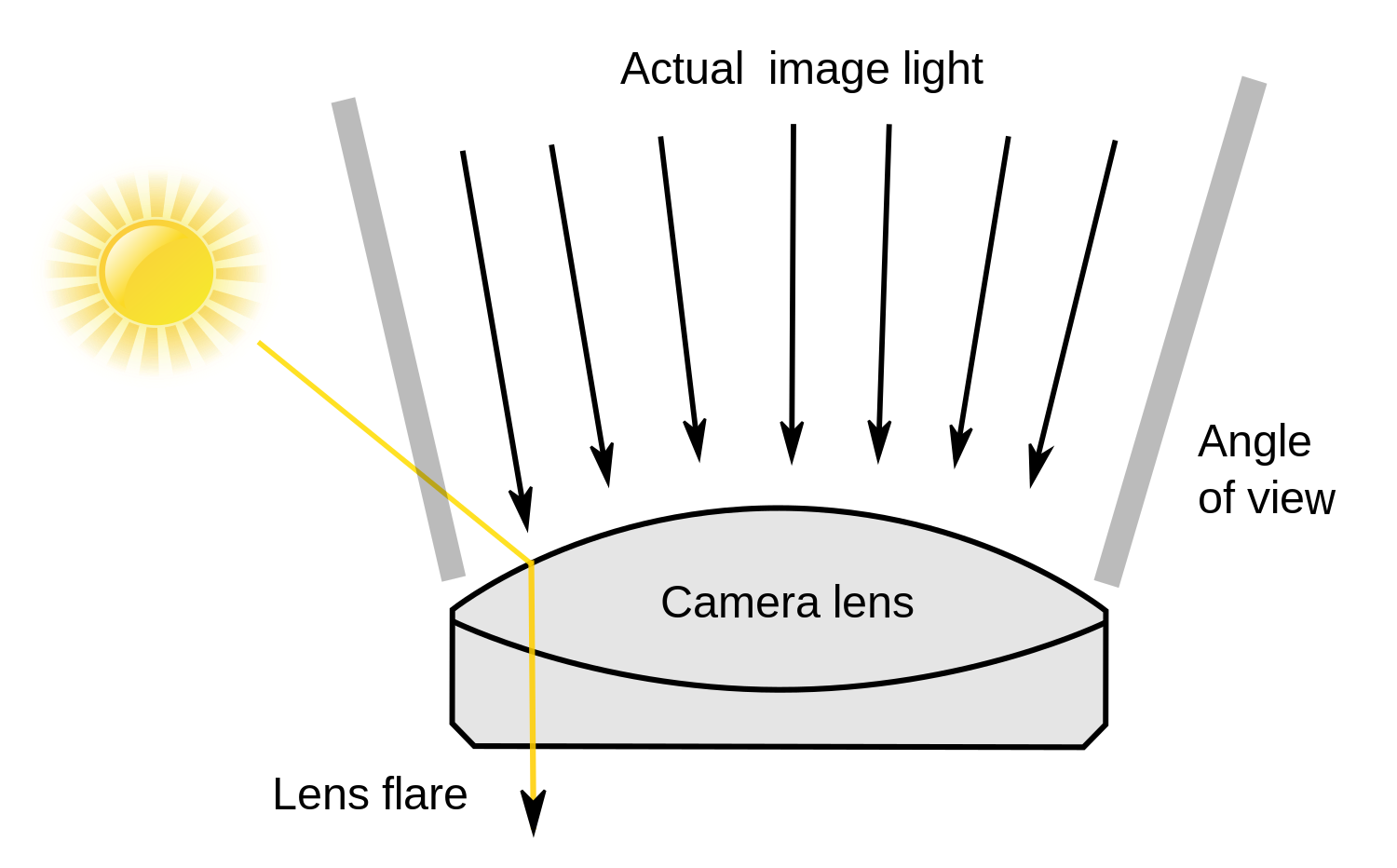}
    \caption{Illustration of how lens hoods help in blocking unwanted light rays that are causing flare artifact.}
    \label{fig:physics of Lens Hoods}
\end{figure}
A lens hood is a simple but essential accessory for cameras and lenses, designed to enhance the quality of photographs. It is a tubular or petal-shaped attachment that is mounted on the front of a camera lens. It blocks and shades the lens from unwanted light sources to capture the intended scene.

A lens hood works by creating a physical barrier around the front of the lens. This barrier prevents light from entering the lens at extreme angles or from off-axis sources. Essentially, it shades the lens from these unwanted light rays as shown in Figure \ref{fig:physics of Lens Hoods}. By doing so, the lens hood helps maintain the intended exposure, enhances color accuracy, and minimizes the risk of lens flare by improving contrast, and reducing glare. Lens hood design is carefully crafted to match the lens's field of view and focal length. Some lens hoods have a round, cylindrical shape, while others have a more distinct petal shape, depending on the lens's aspect ratio and the potential for vignetting (darkening of the corners) in the image. The shape and length of the hood are optimized to effectively block out unwanted light while avoiding any interference with the image frame.

Lens hoods are especially useful in outdoor photography, where you often encounter strong and directional sunlight. They also prove beneficial when shooting in scenarios with artificial light sources. However, it's important to note that not all situations require a lens hood. In some low-light or controlled lighting conditions, using a lens hood may not be necessary and could even become impractical.

\subsection{Limitations}
Hardware-based approaches for mitigating flare, especially in optical systems like cameras and lenses, come with several limitations. Firstly, they can significantly increase the cost of optical systems, which might not be practical for consumer-grade devices. Additionally, the complexity of implementing these solutions, such as specialized lens coatings or the addition of baffles and shades, can make optical systems bulkier and more challenging to manufacture.

Once integrated into a device, these hardware solutions lack flexibility, making it difficult to adapt to changing environmental conditions or different photography scenarios. Furthermore, while aiming to reduce flare, hardware-based solutions may inadvertently affect overall image quality. Striking the right balance between flare reduction and image quality can be a challenge. Moreover, some hardware solutions are specific to certain types of flare and may not be effective against all forms of flare, limiting their applicability. Maintenance and durability are also concerns, as components like lens coatings can wear off over time or be damaged. Therefore, designers and engineers need to carefully consider these limitations and weigh them against the benefits when choosing flare mitigation strategies.

\section{Computational-Based Approaches}
\label{sec:computational-based}
Unlike hardware solutions that often come with limitations and can add complexity and cost, computational methods can be fine-tuned and customized to specific scenarios and environmental conditions. They offer the ability to detect and mitigate flare, making them suitable for dynamic situations. Additionally, computational approaches are implemented in software, reducing the need for additional hardware components and allowing for easy updates and adjustments. The following are some of the computational-based approaches for mitigating the flare effect.

\subsection{Deconvolution}
Deconvolution operation has been used to remove flare, mainly in X-ray imaging \cite{xrayflare1}. Seibert et al. \cite{seibert1985removal} assumed that the removal of veiling glare can be done by knowing the point intensifier spread function (PSF) and applying mathematical deconvolution. Parameters of PSF are represented by $\rho$ and $k$. The parameter $\rho$ indicates the fraction of light strongly scattered in the image intensifier, and $k$ is a measure of the mean travel distance of the scattered photons. Both of those parameters can be determined from a least-square fit of measured contrast ratios versus lead disk diameter from acquired images. However, the derivation and parameterization of the veiling glare PSF in this case is based on the assumption of circular symmetry and spatial invariance of the image intensifier PSF, which is not true for most cases of the flare. It has been shown by experimental results \cite{nalcioglu1984relationship} that the analytical form of the PSF is given as
\begin{equation}
    h(r) = (1-\rho)\frac{\delta(r)}{r} + \frac{\rho}{2kr} e^{-r/k}
\end{equation}

Values of $\rho$ and $k$ can be experimentally determined for the system as discussed before. $\delta(r)$ in the above equation is the Dirac delta function representing a direct mapping term (no veiling glare) which was presented by Paul Dirac in a 1927 paper \cite{DiracThePI}. When those parameters are substituted into Equation 1, they give the system PSF as a function of radial distance where the relation between undergraded image U(r,$\theta$) and the actual detected image V(r,$\theta$) is given by 
\begin{equation}
    V(r,\theta) = [h**U]_{r,\theta}
\end{equation}
Where $**$ represents two-dimensional convolution in the spatial domain, which can be represented in the frequency domain
\begin{equation}
    F_2[V] = F_2[U]F_2[h]
\end{equation}
Where $F_2$ is a 2D Fast Fourier Transform (FFT) operation \cite{fft_1,fft_2, fft_3}. By solving for $F_2[U]$ in Equation 3 and inverse transforming the result
\begin{equation}
    U = F^{-1}_{2}\{F_2[V](\frac{1}{F_2[h]})\}
\end{equation}
Where $F^{-1}_{2}$ is Inverse Fourier Transform. Transforming the PSF is analytically performed by taking advantage of the circular symmetry using the zero-order Bessel function
\begin{equation}
    \hat{H}(f) = 2\pi \int_{0}^{\infty} h(r) J_o(2 \pi r f) r dr 
\end{equation}
Where $\hat{H}$ is the frequency domain representation of h, meaning that:
\begin{equation}
    \hat{H}(f) = F_2[h]
\end{equation}
Substituting Equation 1 into Equation 5, followed by integration and solution, the results are in the frequency domain PSF
\begin{equation}
    \hat{H}(f) = \pi \frac{\rho}{\sqrt{1+(2 \pi k f)^2}} + (1-\rho)
\end{equation}
The inverse frequency filter is obtained by inverting the previous equation
\begin{equation}
    \hat{H}^{-1}(f) = \frac{1}{\pi}\{ \frac{\sqrt{1+(2 \pi k f)^2}}{\rho + (1-\rho)\sqrt{1+(2 \pi k f)^2}} \}
\end{equation}
By multiplying the filter on a point-by-point basis with the frequency domain degraded image, it gives us a product having lower frequencies attenuated. An inverse 2D FFT and image scaling factors are applied to produce an approximate estimate of the original undergraded image in the spatial domain.

\subsection{Automated Lens Flare Removal}
\begin{algorithm}[t]
\caption{Exemplar-based Inpainting \cite{inpainting}.}
\begin{algorithmic} 
\STATE Extract the manually selected initial front $\delta \ohm^0$
\WHILE{Not Done}
\STATE Identify the fill front $\delta \ohm^0$
\IF{$\ohm^t = \phi$}
\STATE EXIT
\ENDIF
\STATE Computer priorities $P(p)$ where $ \forall p \in \delta \ohm^t$
\STATE Find the patch $\psi_{\hat{p}}$ with the maximum priority ($\hat{p} = arg \, max_{p \in \delta \ohm^t} P(p)$)
\STATE Find the exemplar $\psi_{\hat{q}} \in \phi$ that minimizes $d(\psi_{\hat{p}},\psi_{\hat{q}})$
\STATE Copy image data from $\psi_{\hat{q}}$ to $\psi_{\hat{p}}$ $\forall p \in \psi_{\hat{p}} \cap \ohm$
\STATE Update C(p) $\forall p \in \psi_{\hat{p}} \cap \ohm$
\ENDWHILE
\label{algorithm: exemplar_based}
\end{algorithmic}
\end{algorithm}
 Floris Chabert \cite{reflective1} introduced automated detection of the flares using a single input image, particularly reflective flare. This involved a custom blob detection algorithm based on a concept used in OpenCV \cite{opencv} tuned for the specific lens flare that we want to detect and a hybrid inpainting method called exemplar-based inpainting \cite{inpainting}.

The first stage, which is the detection algorithm, includes 5 steps:

\textbf{Multiple Thresholding: }The image is converted to grayscale and binarized using a range of thresholds \cite{thresholding, thresholding_techniques}.

\textbf{Contour Detection: }For each binary image, we then find the contours using a border following the method in \cite{SUZUKI198532}.

\textbf{Blob Merging: }The center of each blob is then computed and blobs from the different binary images are merged depending on their distance and similarity. We finally obtained a set of potential flare candidates.

\textbf{Flare Candidates Filtering: }The flare candidates are pruned using various metrics whose parameters have been tuned using a set of images so as to be robust while avoiding false positives. Those metrics include circularity of the blob, convexity, inertia, and area.

\textbf{Flare Mask Computation: }Finally the mask selecting the flares is computed for the next step

For the second stage, which is recovery using exemplar-based inpainting whose pseudocode is shown in Algorithm 1, a window around the flare is selected, then the following algorithm is executed until all the pixels have been recovered:

\textbf{Identify Fill Front: }We first find the contour of the region we want to fill.

\textbf{Identify Priority Patches: }Patches on the fill front are assigned priorities as to privilege patches that continue strong edges and are surrounded by high-confidence pixels.

\textbf{Find Best Exemplar: }By priority order, we then search the window for known patches that minimize the error.

\textbf{Fill Region using Exemplar Patch: }We finally select pixels from the best patch to fill the masked pixels in the current patch to recover.

This approach is only limited to specific types of flare making it hard to generalize for different type of flare that appears everywhere. In addition to this, it can mistaken any saturated blob and mark it as a flare.
% \subsection{Automatic Flare Spot Artifact Detection and Removal in Photographs by Patricia Vitoria and Coloma Ballester}
% \section{Hardware-Software Co-design}
% \subsection{Glare Aware Photography: 4D Ray Sampling for Reducing Glare Effects of Camera Lenses}
\subsection{Limitations}
Computational-based approaches may not be suitable for all scenarios, particularly when dealing with diverse and unpredictable forms of flare. These methods are often tailored to specific types of flare and may struggle to address the wide range of potential artifacts and aberrations that can occur in optical systems since we can't estimate the luminous intensity, chromatic, and mechanical properties of each camera. Moreover, computational approaches can sometimes misinterpret saturated areas in an image as flare, potentially leading to false positives and unintended alterations. In situations where the nature of flare is highly variable and not easily characterized, alternative solutions, such as learning-based approaches, may be more appropriate to ensure accurate and reliable flare mitigation.

\section{Learning-Based Approaches}
\label{sec:learning-based}
\subsection{Overview of Deep Neural Networks Architectures}
\label{sec:overview of deep neural networks}

\subsubsection{\textbf{Convolutional Neural Network}}
\begin{figure}[t]
    \centering
    \includegraphics[width=\linewidth]{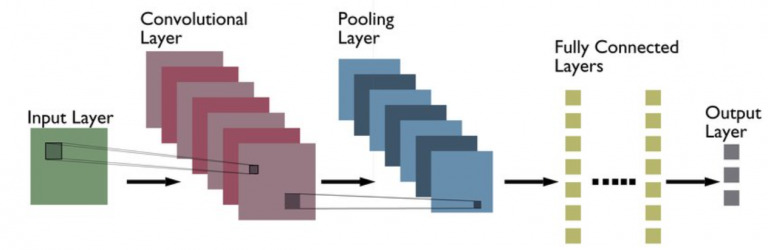}
    \caption{Architecture of convolutional neural networks. From \cite{cnn_arch}.}
    \label{fig:CNN_architecture}
\end{figure}
Convolutional Neural Networks (CNNs) \cite{CNN_intro, cnn_survey_1,cnn_survey_2,cnn_survey_3,cnn_survey_4} stand as one of the most successful and widely adopted architectures within the realm of deep learning, particularly in the field of computer vision. Their inception traces back to Fukushima's groundbreaking work on the Neocognitron \cite{Fukushima1980NeocognitronAS} in 1980, which is a self-organizing neural network model for a mechanism of pattern recognition unaffected by shift in position. This foundational concept paved the way for the evolution of CNNs, eventually leading to their presence in various applications. For instance, Waibel et al. \cite{waibel_et_al} introduced CNNs featuring shared weights among temporal receptive fields and employed backpropagation training for phoneme recognition. Also the first work on modern CNN was by Yann LeCun et al. in their paper “Gradient-Based Learning Applied to Document Recognition” \cite{gradient_cnn} which demonstrated that a CNN model that aggregates simpler features into progressively more complicated features can be successfully used for handwritten character recognition \cite{mnist_dataset_database}. CNNs have revolutionized the field of computer vision by enabling automated feature extraction and hierarchical learning, making them indispensable tools for a wide range of applications.

CNNs typically comprise three primary types of layers as shown in Figure \ref{fig:CNN_architecture}, each serving a distinct purpose. Firstly, there are the convolutional layers where kernels (or filters) are employed to extract features from the input data. Subsequently, nonlinear layers come into play, applying activation functions to the extracted feature maps \cite{activation_function_survey_1,activation_function_survey_2,activation_function_survey_3}. This nonlinearity enables the network to model complex, nonlinear functions effectively. Secondly, pooling layers \cite{pooling_layers_survey} contribute by summarizing information within a small neighborhood of a feature map, often using statistical measures like mean or max, to reduce the spatial resolution. Lastly, the standard neural network \cite{nn_overview_1,nn_overview_2} is employed which is mainly responsible for classification.

One of the distinguishing characteristics of CNNs is their local connectivity, where each unit within a layer receives weighted inputs from a confined region known as the receptive field. By stacking layers to form multi-resolution pyramids, higher-level layers are capable of learning features from increasingly broader receptive fields. This hierarchical approach aids in recognizing complex patterns and structures within the data.

One of the key computational advantages offered by CNNs is weight sharing. In a given layer, all receptive fields share the same set of weights. This ingenious design results in a considerably lower number of parameters compared to normal fully-connected neural networks. Noteworthy CNN architectures that have gained prominence include AlexNet \cite{alexnet}, VGGNet \cite{vgg19}, ResNet \cite{resNet}, GoogLeNet \cite{leNet}, MobileNet \cite{mobileNet}, and DenseNet \cite{densenet}.

\subsubsection{\textbf{Encoder-Decoder Architecture}}
\begin{figure}[t]
    \centering
    \includegraphics[width=\linewidth]{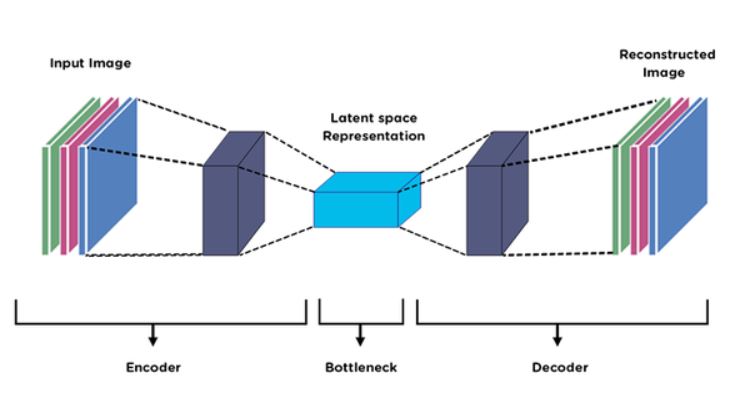}
    \caption{Architecture of a simple encoder-decoder model. From \cite{encoder_decoder_article}}
    \label{fig:encoder_decoder_architecture}
\end{figure}
Encoder-decoder models represent a family of architectures designed to map data from an input domain to an output domain using a two-stage transformation. This process involves two key components: the encoder and the decoder as shown in Figure \ref{fig:encoder_decoder_architecture}. The encoder, denoted as an encoding function, transforms input data into a compressed, latent-space representation, usually named 'z'. This serves as a feature vector capturing essential semantic information. In simpler terms, it extracts meaningful features from the input that are crucial for predicting the output. On the other hand, the decoder learns using this representation to predict the output. This enables efficient information extraction and transfer. These models find extensive application in tasks like image-to-image translation where you need to map one image from one domain to another one (e.g., daytime to nighttime). Also, sequence-to-sequence tasks like machine translation in natural language processing (NLP) can benefit a lot from this architecture. Badrinarayanan et al. introduced SegNet \cite{encoder_decoder} which is based on a deep convolutional encoder-decoder architecture for image segmentation which has proven its accuracy at that time. 

Encoder-decoder models are frequently trained by minimizing a reconstruction loss, typically denoted as $L(y,\hat{y})$ This loss function quantifies the dissimilarities between the ground-truth output $y$ and the reconstructed output $\hat{y}$. Depending on the specific task, the output can take various forms. For instance, it might represent an enhanced version of an image such as image deraining \cite{deraining_survey}, image dehazing \cite{dehazing_survey}, image denoising \cite{denoising_survey}, or super-resolution \cite{super_resolution_survey}. It can also be a segmentation map that highlights different regions within an image.

There is a variation for this architecture named \textbf{Auto-encoders} \cite{bank2021autoencoders} where the input and output are identical, meaning they aim to reconstruct the input itself. This makes them particularly useful for tasks like feature learning, dimensionality reduction \cite{sorzano2014survey}, and anomaly detection \cite{anomaly_detection_survey} due to their ability to learn meaningful representations from data.
\subsubsection{\textbf{Generative Adversarial Networks (GANs)}}
\begin{figure}[t]
    \centering
    \includegraphics[width=\linewidth]{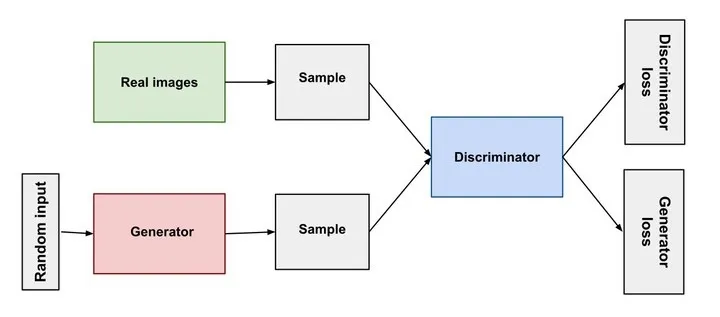}
    \caption{Architecture of a Generative Adversarial Network. From \cite{GANs_architecture}.}
    \label{fig:GANs_architecture}
\end{figure}
\begin{figure}[t]
    \centering
    \includegraphics[width=\linewidth]{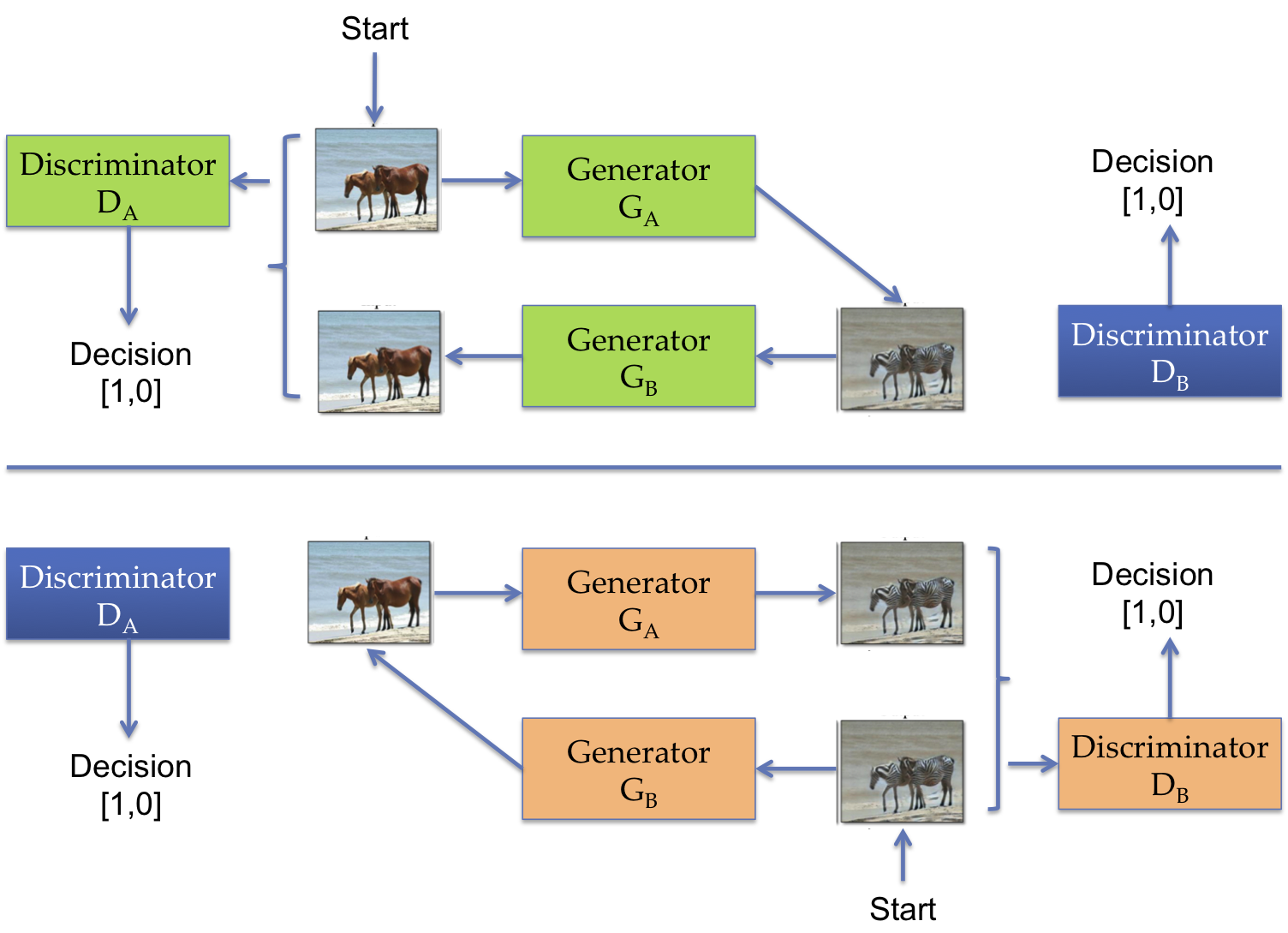}
    \caption{Architecture of CycleGAN. From \cite{cycleGAN_figure}.}
    \label{fig:cycleGANs}
\end{figure}
Generative Adversarial Networks \cite{goodfellow2014generative}, or GANs for short, are a class of deep learning models used in unsupervised machine learning. They were introduced by Ian Goodfellow and his colleagues in 2014. GANs consist of two neural networks, the generator, and the discriminator, that are trained together through adversarial training as shown in Figure \ref{fig:GANs_architecture}. The primary goal of GANs is to generate new data samples that are similar to a given dataset. GANs have been used for a wide range of applications, including image generation, style transfer, and data augmentation.

It can be represented by two networks. The first one is the generator network that takes a random noise vector $z$ as input and maps it to a data sample $x$ that should resemble the real data distribution.
\[x = G(z)\]

The second is the discriminator network, on the other hand, it aims to distinguish between real data samples ($x$) and generated data samples ($G(z)$) where $D(x)$ represents the probability that $x$ is a real data sample.

The training process involves a minimax game, where the generator aims to maximize the probability that the discriminator misclassifies the generated samples, while the discriminator aims to correctly classify real and generated samples. The objective function for GANs can be mathematically defined as:

\begin{equation}
    \min_G \max_D \mathbb{E}_{x\sim p_{\text{data}}(x)} [\log D(x)] + \mathbb{E}_{z\sim p_z(z)} [\log(1 - D(G(z)))]
\end{equation}

Where:

\begin{itemize}
  \item $G$ represents the generator network.
  \item $D$ represents the discriminator network.
  \item $x$ represents real data samples.
  \item $z$ represents random noise vectors sampled from a prior distribution $p_z(z)$.
  \item $\mathbb{E}$ represents the expectation operator.
  \item $p_{\text{data}}(x)$ represents the distribution of real data.
\end{itemize}
In practice, the previous objective function may not provide enough gradient for effectively training the generator especially initially when D can easily discriminate fake samples from real ones. Instead of minimizing $\mathbb{E}_{z\sim p_z(z)} [\log(1 - D(G(z)))]$, we can maximize $\mathbb{E}_{z\sim p_z(z)} [\log(D(G(z)))]$. So now, the new objective function is as follows

\begin{equation}
    \max_G \max_D \mathbb{E}_{x\sim p_{\text{data}}(x)} [\log D(x)] + \mathbb{E}_{z\sim p_z(z)} [\log(D(G(z)))]
\end{equation}

There is another variation of GANs which is \textbf{CycleGAN} \cite{cycleGAN}. It is mainly designed for unpaired image-to-image translation. It was introduced by Zhu et al. in 2017. Unlike traditional GANs that require paired training data (e.g., the input image and corresponding target image), CycleGAN can learn translations between two domains using unpaired data. This makes it particularly useful for tasks like style transfer \cite{9724900}, where there is no one-to-one correspondence between input and target data.

CycleGAN also uses generator-discriminator architecture. However, as shown in Figure \ref{fig:cycleGANs}, it uses two generators and two discriminators:

1- Generator $G_A$ for Domain $A$. It takes an image from domain $B$ and generates a corresponding image in domain $A$.
\[A = G_A(B)\]

2- Generator $G_B$ for Domain $B$. It takes an image from domain $A$ and generates a corresponding image in domain $B$.
\[B = G_B(A)\]

3- Discriminator $D_A$. It distinguishes between real images from domain $A$ and generated images $G_A(B)$. where $D_A(X)$ represents the probability that $X$ is a real image from domain $A$.

4- Discriminator $D_B$. It distinguishes between real images from domain $B$ and generated images $G_B(A)$ where $D_B(X)$ represents the probability that $X$ is a real image from domain $B$.

The objective functions for CycleGAN involve both adversarial losses and cycle consistency losses, ensuring that the generated images not only fool the discriminators but also map back to the original domain correctly.

Cycle-consistency loss for $A \rightarrow B \rightarrow A$:
\begin{equation}
    L_{\text{cycle}}(A, G_A, G_B) = \mathbb{E}_{x\sim p_{\text{data}}(A)} [||G_A(G_B(A)) - A||_1]
\end{equation}

Cycle-consistency loss for $B \rightarrow A \rightarrow B$:
\begin{equation}
    L_{\text{cycle}}(B, G_B, G_A) = \mathbb{E}_{y\sim p_{\text{data}}(B)} [||G_B(G_A(B)) - B||_1]
\end{equation}

Adversarial losses for $A$ and $B$ can be defined similarly to the traditional GAN loss as in Equation 1 or Equation 2.

\subsubsection{\textbf{U-Net}}
\begin{figure}[t]
    \centering
    \includegraphics[width=\linewidth]{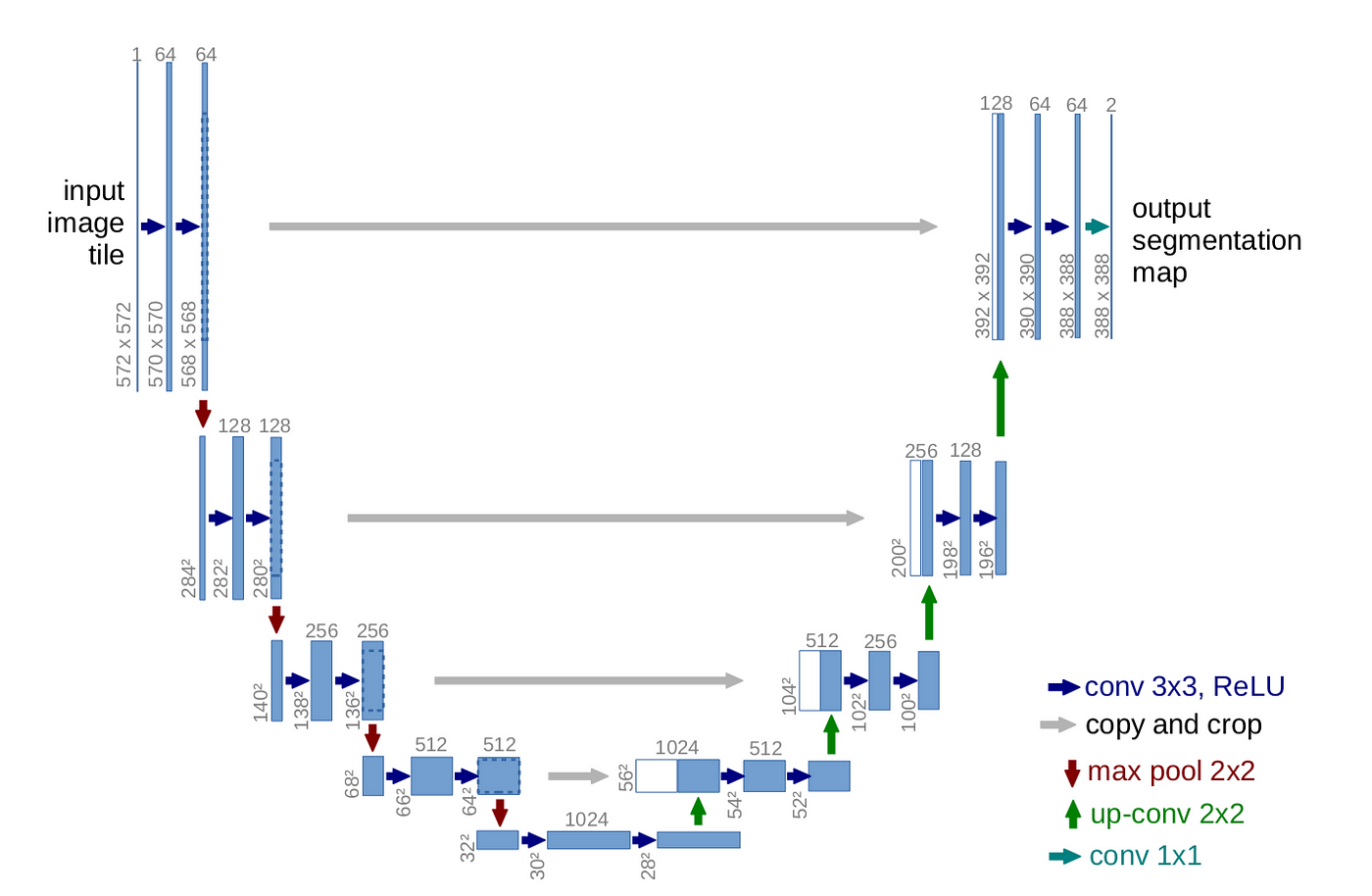}
    \caption{Architecture of U-Net. From \cite{unet_figure}.}
    \label{fig:unet_architecture}
\end{figure}
\begin{figure*}[t]
    \centering
    \includegraphics[width=\linewidth]{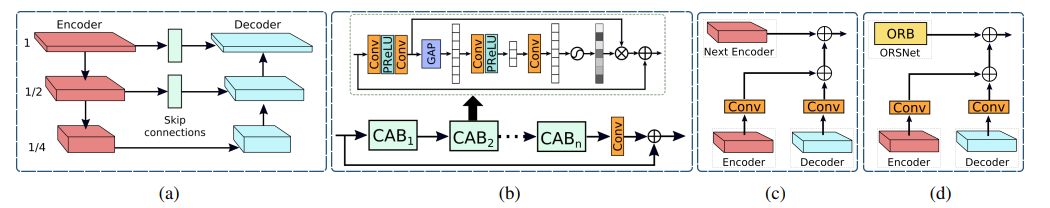}
    \caption{(a) Encoder-decoder subnetwork in MPRNet. (b) Illustration of the original resolution block (ORB) in the original-resolution subnetwork. Each ORB contains multiple channel attention blocks. GAP represents global average pooling \cite{global_average_pooling}. (c) Cross-stage feature fusion between stage 1 and stage 2. (d) Cross-stage feature fusion between stage 2 and the last stage. From \cite{MPRNet}.}
    \label{fig:mpr_overall_architecture}
\end{figure*}
U-Net \cite{unet} is a popular neural network architecture widely used in image processing tasks, especially in image restoration and image segmentation. It was mainly introduced by Olaf Ronneberger et al. for biomedical image segmentation. It is known for its effectiveness in these domains due to its unique design, which resembles the letter "U", hence the name U-Net.

In image restoration tasks, like denoising, deblurring, flare removal, or super-resolution, U-Net operates by taking a noisy or degraded image as input and aims to produce a clean and enhanced version as output. It accomplishes this by employing a contracting path (encoder) on one side of the "U" and an expansive path (decoder) on the other as shown in Figure \ref{fig:unet_architecture}.

The contracting path is responsible for capturing contextual information. It involves a series of convolutional layers followed by downsampling operations (typically max-pooling) that progressively reduce the spatial resolution of the input image. This path enables the network to learn high-level features and understand the global context of the image.

The expansive path, on the other hand, is tasked with restoring fine-grained details. It consists of upsampling layers, which increase the spatial resolution, and convolutional layers which help refine the output. Importantly, this path also incorporates skip connections \cite{resNet}, which connect layers from the contracting path to corresponding layers in the expansive path. These skip connections are a key feature of U-Net and allow the network to access detailed information from earlier stages of processing. This helps in reconstructing complex textures and features that might be lost during downsampling.

The intuition behind U-Net lies in its ability to capture both high-level context information and fine-grained details in images. The contracting path effectively learns to recognize global image features and object boundaries. The expansive path, with the help of skip connections, then refines the output while preserving the spatial details. This combination of capturing context and details is pivotal in achieving accurate pixel-wise operations whether restoration or segmentation. This is crucial for tasks where preserving the structure of objects in the image is essential.

\subsubsection{\textbf{MPRNet}}
\begin{figure}[t]
    \centering
    \includegraphics[width=\linewidth]{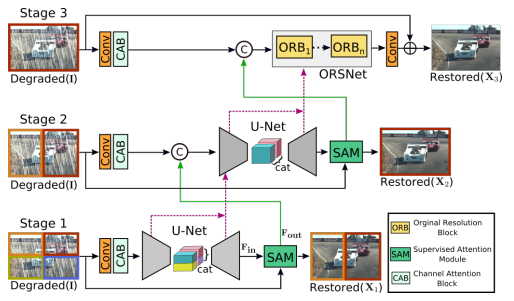}
    \caption{Architecture of multi-stage architecture for progressive image restoration. Earlier stages employ encoder-decoders to extract multi-scale contextualized features, while the last stage operates at the original image resolution to generate spatially accurate outputs. A supervised attention module is added between every two stages that learn to refine features of one stage before passing them to the next stage. Dotted pink arrows represent the cross-stage feature fusion mechanism. From \cite{MPRNet}.}
    \label{fig:mpr_architecture}
\end{figure}
\begin{figure}[t]
    \centering
    \includegraphics[width=\linewidth]{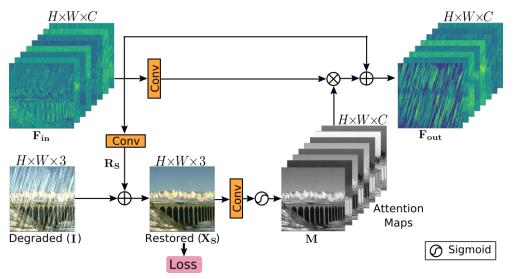}
    \caption{Supervised attention module employed by MPRNet. From \cite{MPRNet}.}
    \label{fig:mpr_attention_module}
\end{figure}

Multi-Stage Progressive Restoration Network \cite{MPRNet} is a proposed deep learning network for image restoration that comprises three stages that work sequentially to progressively enhance images. The first two stages employ encoder-decoder subnetworks based on the standard U-Net \cite{unet}, enabling them to grasp broad contextual information through large receptive fields. MPRNet also employ channel attention blocks (CABs) \cite{CAB_block} to extract features at each scale. The feature maps at U-Net skip connections are also processed with the CAB. Finally, instead of using transposed convolution \cite{transposed_conv} for increasing the spatial resolution of features in the decoder, bilinear upsampling is used followed by a convolution layer. This helps reduce checkerboard artifacts \cite{transposed_conv} in the output image that often arises due to the transposed convolution.
\begin{figure*}[t]
    \centering
    \includegraphics[width=\linewidth]{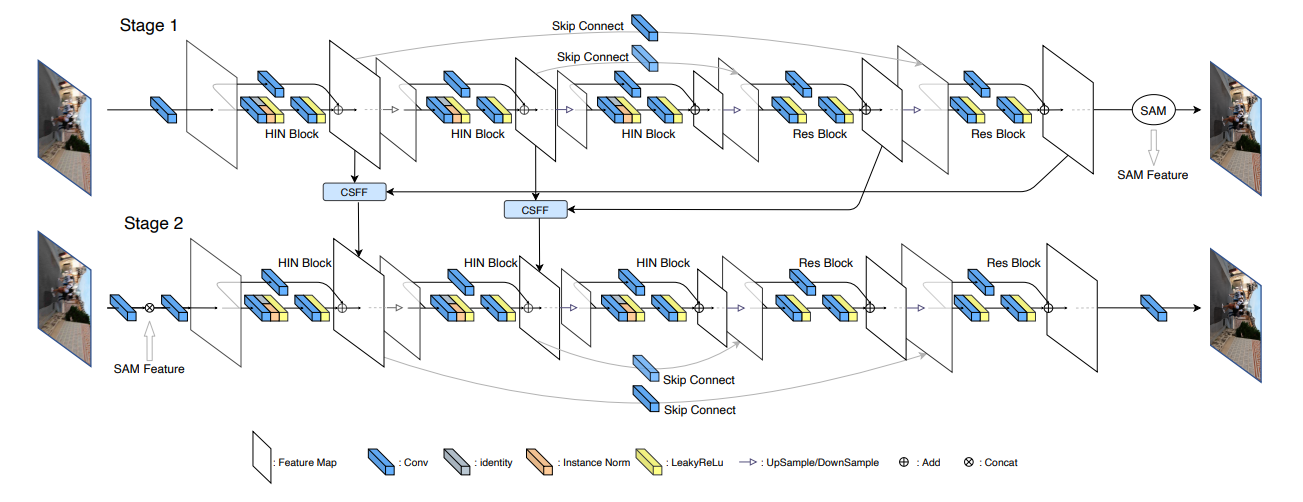}
    \caption{Architecture of Half Instance Normalization Network \cite{HINET}. The encoder of each subnetwork contains Half Instance Normalization Blocks (HIN Block). For simplicity, only 3 layers of HIN Block are shown in the figure, and HINet has a total of 5 layers. CSFF and SAM modules are adopted as in MPRNet \cite{MPRNet}. From \cite{HINET}.}
    \label{fig:HInet_architecture}
\end{figure*}

It's important to note that image restoration is a task that demands precise pixel-to-pixel correspondence between the input and output images. Therefore, the final stage utilizes a subnetwork that operates at the original input image resolution, ensuring the preservation of fine details in the output image.

Rather than simply stacking multiple stages, MPRNet incorporates a supervised attention module between each pair of consecutive stages as shown in Figure \ref{fig:mpr_attention_module}. This module plays a crucial role in rescaling the feature maps from the previous stage based on the guidance of ground-truth images before passing them to the subsequent stage. Additionally, it introduces a cross-stage feature fusion mechanism that leverages the intermediate multi-scale contextualized features from earlier subnetworks to consolidate the intermediate features of the latter subnetwork. This cross-stage feature fusion is illustrated in Figure \ref{fig:mpr_overall_architecture} (c) and (d). In order to preserve fine details from the input image to the output image, the original-resolution subnetwork (ORSNet) in the last stage is used where it does not employ any downsampling operation and generates spatially enriched high-resolution features. It consists of multiple original resolution blocks (ORBs) as shown in Figure \ref{fig:mpr_overall_architecture} (b), each of which further contains CABs.

While MPRNet comprises multiple stages as shown in Figure \ref{fig:mpr_architecture}, it's worth noting that each stage has access to the input image. Following the approach of recent restoration methods \cite{image_restoration1, image_restoration2}, they implemented a multi-patch hierarchy on the input image, splitting it into non-overlapping patches: four for stage 1, two for stage 2, and retaining the original image for the last stage.
\begin{figure*}[t]
    \centering
    \includegraphics[width=\linewidth]{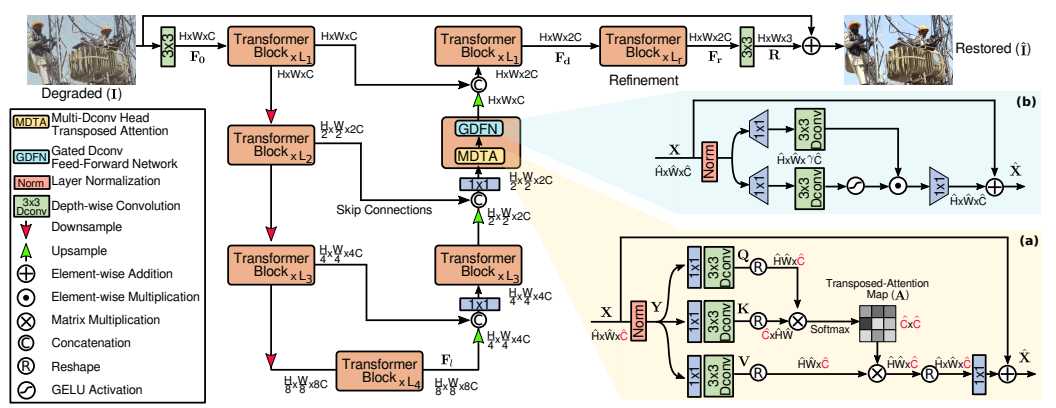}
    \caption{Architecture of Restormer for high-resolution image restoration. Restormer consists of a multi-scale hierarchical design incorporating efficient Transformer blocks. The core modules of the Transformer block are: (a) multi-Dconv head transposed attention (MDTA) that performs (spatially enriched) query-key feature interaction across channels rather than the spatial dimension, and (b) Gated-Dconv feed-forward network (GDFN) that performs controlled feature transformation, i.e., to allow useful information to propagate further. From \cite{Restormer}.}
    \label{fig:Restormer Architecture}
\end{figure*}

At any given stage S, instead of directly predicting a fully restored image $X_S$, MPRNet predicts a residual image $R_S$. This residual image $R_S$ is then added to the degraded input image I to obtain the final restored image: $X_S = I + R_S$. We can optimize MPRNet end-to-end using the following loss function:

\begin{equation}
L = \sum_{S=1}^{3} [L_{\text{char}}(X_S, Y) + \lambda L_{\text{edge}}(X_S, Y)],
\end{equation}

Here, Y represents the ground-truth image, and we utilize the Charbonnier loss \cite{Charbonnier1994TwoDH}, denoted as \(L_{\text{char}}\), defined as:

\begin{equation}
L_{\text{char}} = \sqrt{(X_S - Y)^2 + \epsilon^2},
\end{equation}

For practical reasons, usually \(\epsilon\) is set to \(10^{-3}\) for all experiments. Additionally, edge loss is used, \(L_{\text{edge}}\), defined as:

\begin{equation}
L_{\text{edge}} = \sqrt{(\Delta(X_S) - \Delta(Y))^2 + \epsilon^2},
\end{equation}

where \(\Delta\) represents the Laplacian operator. The parameter \(\lambda\) in the loss function controls the relative importance of the two loss terms and is set to 0.05, as in previous work (Multi-Scale Progressive Fusion Network for Single Image Deraining \cite{DBLP:journals/corr/abs-2003-10985}).
\begin{figure}[t]
    \centering
    \includegraphics[width=\linewidth]{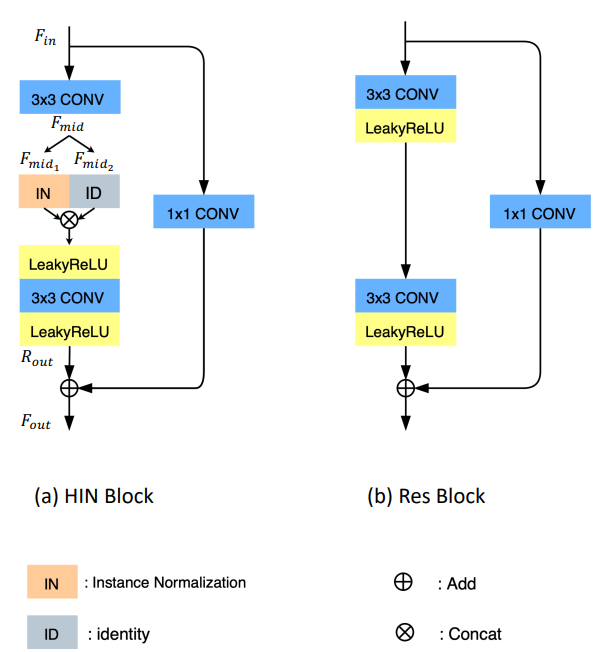}
    \caption{HINet's Half Instance Normalization Block (HIN Block) and ResBlock in details. From \cite{HINET}.}
    \label{fig:HIblock_architecture}
\end{figure}

\subsubsection{\textbf{HINet}} 
Half Instance Normalization Network \cite{HINET} consists of two subnetworks, each of which is based on U-Net \cite{unet} as shown in Figure \ref{fig:HInet_architecture}. In the encoder component, they designed Half Instance Normalization Blocks to extract features in each scale and double the channels of features when downsampling.  In the decoder component, they used ResBlocks \cite{resNet} to extract high-level features, and fuse features from the encoder component to compensate for the loss of information caused by resampling. As for ResBlock, leaky ReLU \cite{leaky_relu} was used with a negative slope equal to 0.2, and batch normalization was removed.

To connect the two subnetworks, the cross-stage feature fusion (CSFF) module and supervised attention module (SAM) were used, these two modules were introduced by Zamir et al. \cite{zamir2020cycleisp} and are the same ones used in MPRNet \cite{MPRNet}. As for the CSFF module, a 3 × 3 convolution was used to transform the features from one stage to the next stage for aggregation, which helps to enrich the multi-scale features of the next stage. As for SAM, the 1 × 1 convolutions in the original module were replaced with 3 × 3 convolutions, and bias was added in each convolution. By introducing SAM, the useful features at the current stage can propagate to the next stage and the less informative ones will be suppressed by the attention masks. Then we optimize HINet end-to-end as follows:
\begin{equation}
\text{Loss} = -\sum_{i=1}^{2} \text{PSNR}((R_i + X_i), Y)\footnote{PSNR metric is explained in details in Section \ref{sec:psnr}}
\end{equation}
Where
\begin{itemize}
    \item $X_i \in \mathbb{R}^{N \times C \times H \times W}$ denote the input of subnetwork $i$, where $N$ is the batch size of data, $C$ is the number of channels, and $H$ and $W$ are spatial dimensions.
    \item $R_i \in \mathbb{R}^{N \times C \times H \times W}$ denotes the final prediction of subnetwork $i$.
    \item $Y \in \mathbb{R}^{N \times C \times H \times W}$ is the ground truth in each stage.
\end{itemize}
Because of variance of small image patches differ a lot among mini-batches and the different formulations of training and testing \cite{wide_activation}, Batch Normalization (BN) \cite{batch_norm} is not commonly used in low-level tasks as deraining \cite{deraining}, deblurring \cite{deblurring}, and super-resolution \cite{super_resolution}. Instead, Instance Normalization (IN) \cite{instance_norm} is used which keeps the same normalization procedure consistent in both training and inference. Further, IN re-calibrates the mean and variance of features without the influence of batch dimension, which can keep more scale information than BN. IN is used to build Half Instance Normalization Block (HIN block). By introducing the HIN block, the modeling capacity of HINet is improved. Moreover, the extra parameters and computational costs introduced by IN can be ignored. The complete structure of the HIN block can be shown in Figure \ref{fig:HIblock_architecture}.
\begin{figure*}[t]
    \centering
    \includegraphics[width=\linewidth]{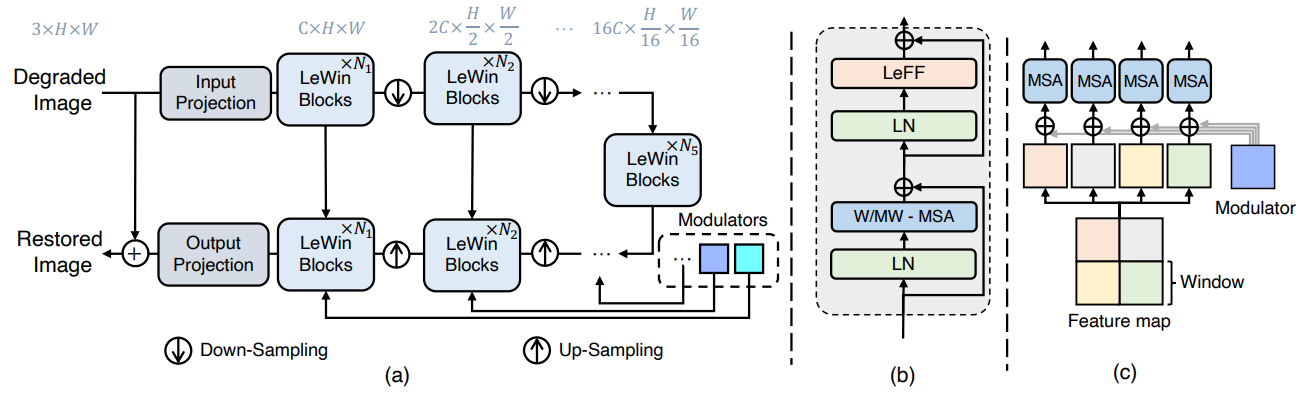}
    \caption{ (a) Overview of the Uformer structure. (b) LeWin Transformer block. (c) Illustration of how the modulators modulate the W-MSAs in each LeWin Transformer block which is named MW-MSA in (b). From \cite{uformer}.}
    \label{fig:Uformer_overall_architecture}
\end{figure*}
\subsubsection{\textbf{Restormer}}
The Restormer \cite{Restormer} architecture is designed for efficient image restoration tasks, especially for handling high-resolution images. It is based on Transformer architecture \cite{attention_is_all_you_need} which was first developed for sequence processing in natural language tasks. It has been adapted in numerous vision tasks such as image recognition \cite{image_reco1, image_restoration2}, segmentation \cite{transformer_segmentation1, transformer_segmentation2, transformer_segmentation3}, object detection \cite{transformer_object1, transformer_object2, transformer_object3}. The Vision Transformers \cite{vision_transformer,vision_transformer_survey, image_recognition2} decompose an image into a sequence of patches (local windows) and learn their mutual relationships. The distinguishing feature of these models is the strong capability to learn long-range dependencies between image patch sequences and adaptability to given input content. Due to these characteristics, Transformer models have also been studied for low-level vision problems. To handle large-resolution images, the researchers for Restormer incorporated key innovations in their multi-scale hierarchical Transformer model.

The Restormer pipeline begins with a degraded input image (I) of dimensions H×W×3. Initially, a convolutional layer is applied to create low-level feature embeddings ($F_0$) with dimensions H×W×C, where C represents the number of channels. These shallow features are then processed through a 4-level symmetric encoder-decoder, resulting in deep features ($F_d$) with dimensions H×W×2C. Each encoder-decoder level contains multiple Transformer blocks, with the number of blocks increasing from top to bottom levels. This hierarchical design allows for efficient spatial size reduction and channel capacity expansion. The decoder takes low-resolution latent features ($F_l$) as input and progressively restores high-resolution representations.

The Restormer model employs pixel-unshuffle and pixel-shuffle operations \cite{pixel_shuffle_unshuffle} for feature downsampling and upsampling, respectively. Skip connections \cite{unet} connect encoder features with decoder features to assist the recovery process. Furthermore, a 1×1 convolution reduces channels at all levels except the top one. Transformer blocks aggregate low-level encoder features with high-level decoder features at level 1, preserving fine structural and textural details in the restored images. Deep features ($F_d$) undergo further refinement at high spatial resolution, contributing to quality improvements.

In the final stage, a convolutional layer is applied to refined features, generating a residual image (R) with dimensions H×W×3. This residual image is added to the degraded input image (I) to obtain the restored image ($\hat{I} = I + R$). The core components of the Restormer Transformer block include the following as shown in Figure \ref{fig:Restormer Architecture}:

\textbf{Multi-Dconv Head Transposed Attention (MDTA):} MDTA addresses the computational complexity of self-attention layers in Transformers \cite{vision_transformer, image_recognition2}. Unlike conventional self-attention, which has quadratic complexity with input resolution, MDTA operates linearly by computing cross-covariance across channels. Depth-wise convolutions emphasize local context before calculating feature covariance, resulting in a global attention map that implicitly encodes global context.

\textbf{Gated-Dconv Feed-Forward Network (GDFN):} GDFN enhances representation learning in the feed-forward network \cite{feed_forward_network}. It incorporates a gating mechanism and depth-wise convolutions \cite{depthwise}. The gating mechanism involves element-wise products of linear transformations, one of which employs the GELU non-linearity \cite{GELU}. Depth-wise convolutions \cite{depthwise} capture information from neighboring pixel positions, aiding in learning local image structures for effective restoration.
\begin{figure}[t]
    \centering
    \includegraphics[width=\linewidth]{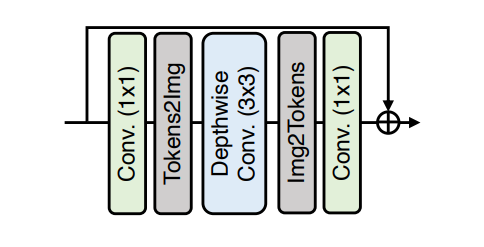}
    \caption{Locally-enhanced feed-forward network employed by the Uformer. From \cite{uformer}.}
    \label{fig:locally_enhanced_feed_forward}
\end{figure}
\begin{figure*}[t]
    \centering
    \includegraphics[width=\linewidth]{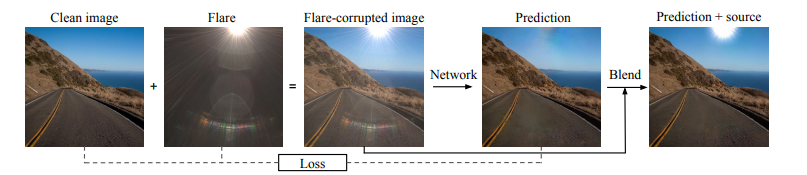}
    \caption{Wu et al.'s approach includes three steps: 1) Training input by composing a flare-free image and a flare image. 2) A CNN-based network was trained to restore the flare-free scene without the light source 3) After prediction, the input light source was blended back into the output image. From \cite{wuetal}.}
    \label{fig:wu et al approach}
\end{figure*}
\textbf{Progressive Learning:} To address the challenge of training on fixed-size image patches, Restormer adopts progressive learning. In early epochs, the model trains on smaller image patches and gradually transitions to larger patches in later epochs. This strategy allows the network to encode global image statistics, improving performance on full-resolution images during testing.
\subsubsection{\textbf{Uformer}}
Uformer \cite{uformer} is a hierarchical network designed to address the challenges of image restoration tasks efficiently. Uformer employs a U-shaped \cite{unet, image_to_image_translation_with_conditional_adversarial_networks} hierarchical network structure with skip connections between the encoder and decoder components. Given an input degraded image (I) with dimensions 3×H×W, Uformer initially applies a 3×3 convolutional layer with LeakyReLU activation \cite{leaky_relu} to extract low-level features ($X_0$) with dimensions C×H×W. It follows a U-shaped architecture, containing K encoder stages, each incorporating LeWin Transformer blocks and a down-sampling layer. These stages hierarchically reduce spatial size while expanding channel capacity. A bottleneck stage at the end of the encoder captures longer dependencies due to the hierarchical structure.

For feature reconstruction, K decoder stages with up-sampling layers and LeWin Transformer blocks, similar to the encoder, are employed. Transposed convolution is used for upsampling \cite{transposed_conv}. Features from the decoder are concatenated with corresponding encoder features through skip-connections \cite{resNet} before input to LeWin Transformer blocks for image restoration. After the K decoder stages, a 3×3 convolutional layer is applied to produce a residual image (R) of dimensions 3×H×W. The restored image is obtained by adding $R$ to the degraded input image ($\hat{I} = I + R$). Uformer is trained using the Charbonnier loss too \cite{Charbonnier1994TwoDH, charbonnier_loss}, which measures the difference between the ground-truth image ($I$) and the restored image ($\hat{I}$).

Uformer faces two significant challenges when applying the Transformer architecture to image restoration. Firstly, standard Transformers compute global self-attention across all tokens, leading to high computational costs for high-resolution feature maps \cite{attention_is_all_you_need, vision_transformer, vision_transformer_survey}. Secondly, local context information is crucial for image restoration, and previous Transformers show limitations in capturing local dependencies \cite{li2021localvit, wu2021cvt}. To address these issues, Uformer introduces the locally enhanced Window (LeWin) Transformer block as shown in Figure \ref{fig:locally_enhanced_feed_forward}. This block combines self-attention for long-range dependencies and convolution for local context. The block consists of two core components: Window-based Multi-head Self-Attention (W-MSA) \cite{transformer_object2, shaw2018selfattention} and Locally-enhanced Feed-Forward Network (LeFF) \cite{sandler2019mobilenetv2, li2021localvit, shaw2018selfattention}. W-MSA performs self-attention within non-overlapping local windows, significantly reducing computational costs compared to global self-attention. LeFF includes depth-wise convolutions \cite{depthwise} to capture local information essential for image restoration.

 Uformer incorporates a multi-scale restoration modulator to enhance its capability to handle various image perturbations. Different types of image degradation, such as blur or noise, have distinct patterns requiring specific restoration approaches. The modulator introduces minimal additional parameters and computational cost. In each LeWin Transformer block of the decoder, a modulator is applied as a learnable tensor with a shape of $M \times M \times C$, where M is the window size, and $C$ is the channel dimension of the current feature map. These modulators act as shared bias terms added to non-overlapping windows before the self-attention module. This lightweight addition allows for flexible feature map adjustments, improving restoration details. The multi-scale restoration modulator is particularly effective in tasks like image deblurring \cite{deblurring_using_transformer} and denoising \cite{denoising_using_transformer}, enhancing restoration results with minimal computational overhead. The complete Uformer architecture is shown in Figure \ref{fig:Uformer_overall_architecture}.

\subsection{How to Train Neural Networks for Flare Removal}
Wu et al. \cite{wuetal} introduced the first deep learning method to remove flare using U-Net \cite{unet}, they modeled the flare as a layer that can be added linearly on top of flare-free images to generate flare-corrupted images where flare-free images are sampled from Flickr24K used in \cite{flickr24k}. The generation of flare images is explained in detail in the Section \ref{sec:flare datasets}. 
\begin{figure*}[t]
    \centering
    \includegraphics[width=\linewidth]{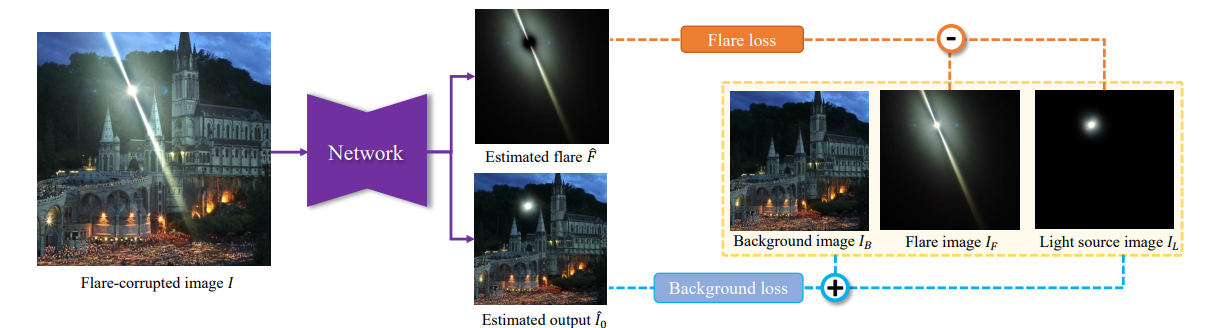}
    \caption{Dai et al.'s approach in Flare7K++ where paired flare and light source images are randomly selected from the Flare7K++ dataset. The flare image will be added to the background image to synthesize the flare-corrupted image. From \cite{flare7kpp}.}
    \label{fig:flare7k and flare7k++}
\end{figure*}

One of the main reasons why flare removal is hard is that if the deep learning model is trained naively, it tends to darken the image and remove the light source as a part of the flare which is not the intended function. To avoid that, they blended the light source with the network's output before computing the loss to avoid penalizing the light source region. The binary saturation mask $M$ for the light source which is blended with the output can be obtained using thresholding on luminance based on the observation that the flare-causing light source is likely saturated in the input image. Then by applying morphological operations, small saturated regions are excluded from the mask. Any pixel inside the mask $M$ will be replaced with the ground truth $I_0$. So now, the predicted flare-free image $\hat{I}_0$ can be computed using the following equation:
\begin{equation}
    \hat{I}_0 = I_0 \odot M + f(I_F,\theta) \odot (1-M)
\end{equation}
Where $f(I_F,\theta)$ represents the output of the network which is U-Net.

To encourage the predicted flare-free $\hat{I}_0$ image to be close to the ground truth $I_0$, they used absolute loss on the RGB values between them. In addition to that, they used perceptual loss by feeding both images into pre-trained VGG-19 like in \cite{flickr24k}. the image loss can be computed using the following equation:
\begin{equation}
    L_I = \Vert \hat{I_0}-I_0 \Vert_1 + \sum_l \lambda_l \Vert \phi_l(\hat{I_0}) - \phi_l(I_0) \Vert_1
\end{equation}

They also wanted to encourage the similarity between predicted flare $\hat{F}$ and the ground truth flare $F$. They applied the same loss as in Equation 2 but to the flare ones where $\hat{F}$ can be obtained by calculating the difference between the network input and the masked network output:
\begin{equation}
    \hat{F} = I_F - f(I_F,\theta) \odot (1-M)
\end{equation}
So now, the total loss is
\begin{equation}
    L = L_I + L_F
\end{equation}
For post-processing during inference only, the mask $M$ is feathered to create a gradual transition at its boundaries to construct $M_f$, and then the network's input and output and be blended using said feathered mask in linear space
\begin{equation}
    I_B = I_F \odot M_f + f(I_f,\theta) \odot (1-M_f)
\end{equation}
where $I_F$ is the flare-corrupted image entering the network. The whole training pipeline can be shown in Figure \ref{fig:wu et al approach}.

\subsection{Flare7K++}
Dai et al. \cite{flare7kpp} introduced a modification to the previous approach. They addressed the limitations of obtaining a light source mask through thresholding, which could fail to accurately segment light sources that are not sufficiently bright. Moreover, in some cases, the streak region might become overexposed and be mistakenly considered as part of the light source. To avoid these issues, they trained the network end-to-end to directly predict a 6-channel output as shown in Figure \ref{fig:flare7k and flare7k++}, with the first three channels as a flare-free image $\hat{I}_0$ and the last three channels as a flare image $\hat{F}$ using the light source annotation which they introduced in Flare7K dataset \cite{flare7k}. Light source image $I_L$ is added to the background images $I_B$ to synthesize the ground truth of flare-free images $I_0$. Then, the difference between the flare image $I_F$ and light source image $I_L$ is calculated to get the ground truth flare image $F$. They tried different network architectures like U-Net \cite{unet}, MPRNet \cite{MPRNet}, HINet \cite{HINET}, Restormer \cite{Restormer}, and Uformer \cite{uformer}.

To linearize each image before addition or subtraction, they applied an inverse gamma correction curve with $\gamma$ sampled from [1.8,2.2]. They encouraged similarity between predicted flare-free image and the ground truth using absolute and perceptual losses
\begin{equation}
    L_I = L_1(\hat{I}_0,I_0) + L_{vgg}(\hat{I}_0,I_0) 
\end{equation}

The same loss is applied to the predicted flare and the actual flare to encourage their similarity too. They also introduced new reconstruction loss to ensure that the predicted flare-free image and predicted flare can be added to get the original input
\begin{equation}
    L_{rec} = | I- Clip(\hat{I}_0,\hat{F}) |
\end{equation}
where $\oplus$ means the addition operation in the linearized gamma-decoded domain with the previously sampled $\gamma$. Then, the addition is clipped to the range of [0,1]. Overall, the final loss function aims to minimize a weighted sum of all these
losses:
\begin{equation}
    L = w_1 L_I + w_2 L_F + w_3 L_{rec}
\end{equation}
where $w_1$, $w_2$, and $w_3$ are respectively set to 0.5, 0.5, and 1.0 in their experiments.

\subsection{Optical Center Symmetry Prior}
\begin{figure*}[t]
    \centering
    \includegraphics[width=\linewidth]{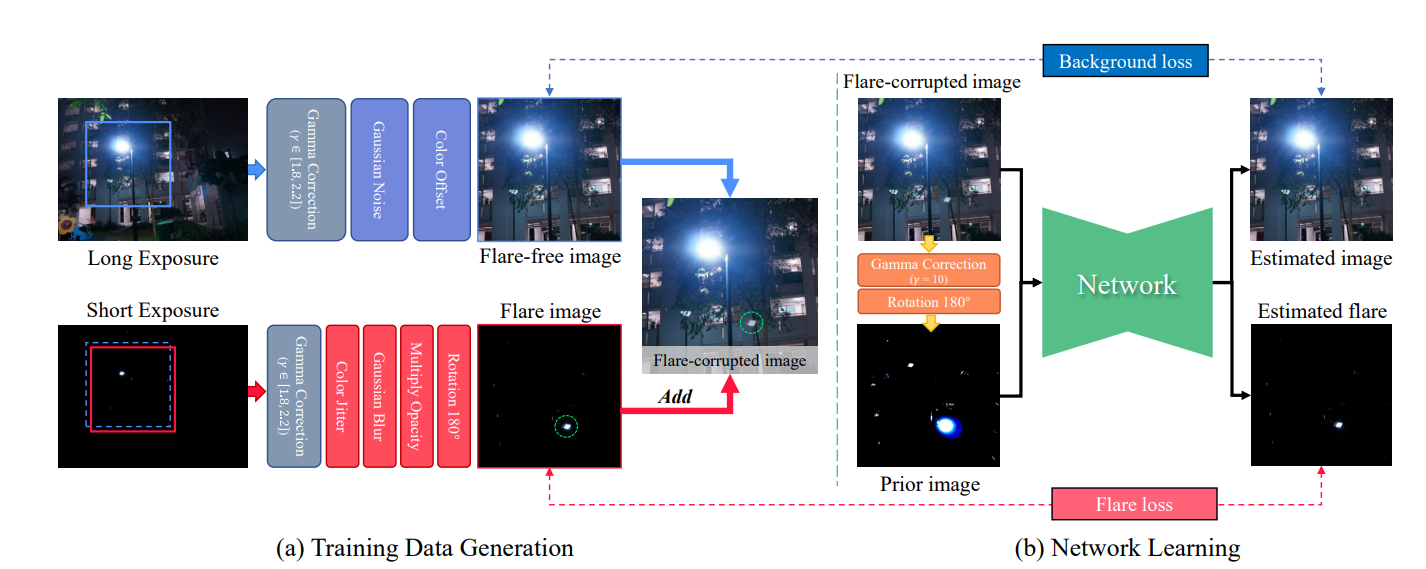}
    \caption{Dai et al.'s training pipeline in Bracket Flare. From \cite{dai2023nighttime}.}
    \label{fig:optical center symmetry prior}
\end{figure*}
Dai et al. \cite{dai2023nighttime} proposed an optical center symmetry prior, which suggests that the reflective flare and light source are always symmetrical around the lens’s optical center. This prior helps to locate the reflective flare’s proposal region more accurately and can be applied to most smartphone cameras.

As the optical center symmetry prior is a global prior, it is difficult for CNN-based networks to learn it, and even Transformer-based networks struggle to learn such symmetry. Therefore, they incorporated this prior by applying a gamma correction of $I_g = I^\gamma$ to the input flare-corrupted image with $\gamma$ = 10. This operation extracts the nearly saturated areas of the light source in each channel. We then rotate this nearly saturated image by 180 degrees to obtain a prior image, which is concatenated with the flare-corrupted image to form a six-channel input. This rotated prior image provides an initial approximation of the reflective flare’s pattern. The entire pipeline for training the network is shown in Figure \ref{fig:optical center symmetry prior}. Obtaining both long and short-exposure shots is explained in Section \ref{sec:flare datasets}.

They adopted different network structures for their experiments, they used the default network settings of MPRNet, HINet, and Uformer and reduced the feature channels of Restormer from 48 to 24 due to GPU memory constraint

For the loss functions, they followed the previous work using absolute and perceptual loss for both the background and the flare image. In addition to this, they also used reconstruction loss as in Flare7K++ but removed the clipping, so the reconstruction loss can be expressed as:
\begin{equation}
    L_{rec} = | I - (\hat{I}_0 \oplus \hat{F})  |
\end{equation}
Where $\hat{I}_B$ and $\hat{I}_F$ are the predicted flare-free image and predicted flare respectively. The main difference from the previous work is the $L_{mask}$. Since the regions of reflective flares are always small, a simple $L_1$ loss may ignore these regions, leading to the local optimum. To encourage the network to focus on restoring the flare-corrupted regions, they calculated the flare’s mask $I_M$ from the ground truth of flare image $I_F$. This masked loss can be written as:

\begin{equation}
    L_{mask} = | I_M * (\hat{I} - I) |
\end{equation}
So now, the background and flare loss can be written as:
\begin{equation}
    L_{I/F} = w_1 L_1 + w_2 L_{vgg} + w_3 L_{mask}
\end{equation}
where $w_1$, $w_2$, and $w_3$ are respectively set to 0.5, 0.1, and 20.0. So now, the total loss is: 
\begin{equation}
    L = L_{I} + L_F + L_{rec}
\end{equation}

\section{Flare Datasets}
\label{sec:flare datasets}
Data availability has always been the major concern for most of the ground-breaking research in the field of flare removal due to the lack of paired flare-corrupted and flare-free images. Because constructing paired images is a labor-intensive task and can't be 100\% reliable and accurate due to image alignment issues, most of the researches focused on synthetic and semi-synthetic flare generation. This survey not only covers the used datasets but also how they were created.
\begin{figure}[t]
    \centering
    \begin{minipage}[b]{.48\linewidth}
      \centering
      \includegraphics[width=4.0cm]{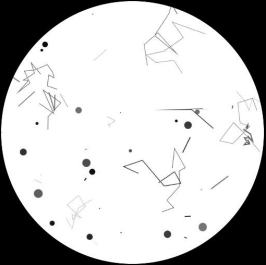}
      \centerline{(a) Aperture of dirty lens}\medskip
    \end{minipage}
    \hfill
    \begin{minipage}[b]{0.48\linewidth}
      \centering
      \centerline{\includegraphics[width=4.0cm]{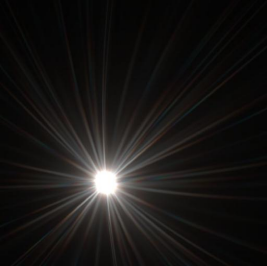}}
      \centerline{(b) Simulated Flare}\medskip
    \end{minipage}
    \caption{To simulate a scattering flare component, they sampled a number of apertures with random dots and lines that resemble defects as in (a). Wave optics then can be used to compute the flare image (b) of any light source imaged by that synthetic aperture.}
    \label{fig:wu_scattering}
\end{figure}

\begin{figure}[t]
    \centering
    \includegraphics[width=\linewidth]{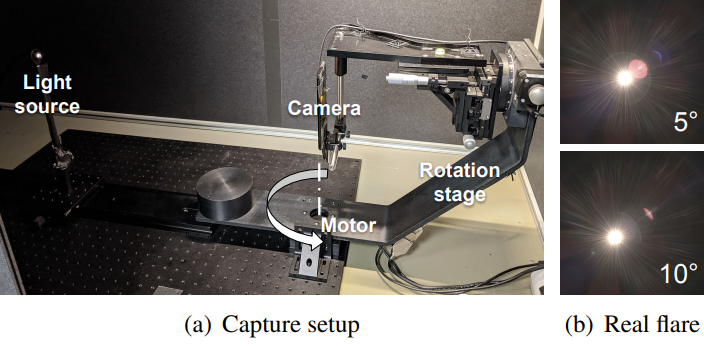}
    \caption{Wu et al.'s setup for capturing images of real lens flare using a strong light source in a dark room using a motorized rotation stage that reproduces a wide range of incident angles. From \cite{wuetal}.}
    \label{fig:wu_reflective}
\end{figure}

\subsection{Wu el al}
Wu et al. \cite{wuetal} took a major step toward making the flare removal dataset available where they noticed that the additive nature of light implies that we can model flare as an additive artifact on top of the flare-free image. So they tried to synthesize flare using physics-based methods to model and generate the flare image so that they could add it to flare-free images to generate flare-corrupted images. \textbf{For the scattering flare}, they assumed that under the thin-lens approximation, an optical imaging system can be characterized by the complex-valued pupil function $P(u, v)$: a 2D field describing, for each point $(u, v)$ on the aperture plane, the lens's effect on the amplitude and phase of an incident wave with wavelength $\lambda$:
\begin{equation}
     P_\lambda(u, v) = A(u, v) \cdot \exp (i\phi_\lambda(u, v)) .
\end{equation}

Here, \( A \) is an aperture function, a property of the optics that represents its attenuation of the incident wave's amplitude.\footnote{Strictly speaking, lens optics can also introduce a phase shift, in which case A becomes a complex-valued function. However, this has shown little difference in their simulation results, so they assumed A is real-valued.} In its simplest form, a camera with an aperture of a finite radius \( r \) has an aperture function of
\begin{equation}
A(u, v) = \begin{cases}
1 & \text{if } u^2 + v^2 < r^2 \\
0 & \text{otherwise}
\end{cases} . 
\end{equation}

\(\phi_\lambda\) in Eq. 1 describes the phase shift, which depends on the wavelength as well as the 3D location of the light source$(x, y, z)$. Omitting the aperture coordinates$(u, v)$, \(\phi_\lambda\) can be written as:

\begin{equation}
    \phi_\lambda(x, y, z) = \phi^S_\lambda(x/z, y/z) + \phi^{DF}_\lambda(z)
\end{equation}

where the linear term $\phi_{\lambda}^{S}$ is determined by the angle of incidence, and the defocus term $\phi^{DF}_{\lambda}$ depends on the depth z of the point light source. Once fully specified, the pupil function P in Eq. 1 can be used to calculate the point spread function (PSF) by a Fourier transform (F) \cite{fourier}:
\begin{equation}
PSF_\lambda = |F\{P_\lambda\}|^2
\end{equation}
\begin{figure*}[ht]
    \centering
    \includegraphics[width=\linewidth]{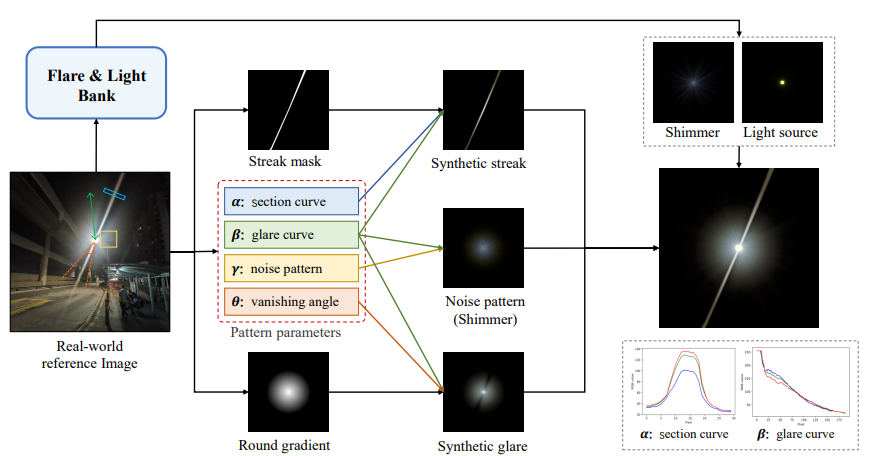}
    \caption{The pipeline for scattering flare synthesis in Flare7K \cite{flare7k}. To synthesize scattering flare, they first obtained streak section curve $\alpha$, glare descent curve $\beta$, noise patch near the light source $\gamma$, and the vanishing corner’s angle $\theta$ around the streak from the reference image. $\alpha$ and $\theta$ are used to synthesize the glare effect while $\alpha$ and $\beta$ are used to simulate the streak. To simulate the degradation around the light source, they added a blurred fractal noise pattern on the shimmer to create a realistic flare. From \cite{flare7kpp}.}
    \label{fig:flare7k}
\end{figure*}
\begin{figure*}[ht]
    \centering
    \includegraphics[width=\linewidth]{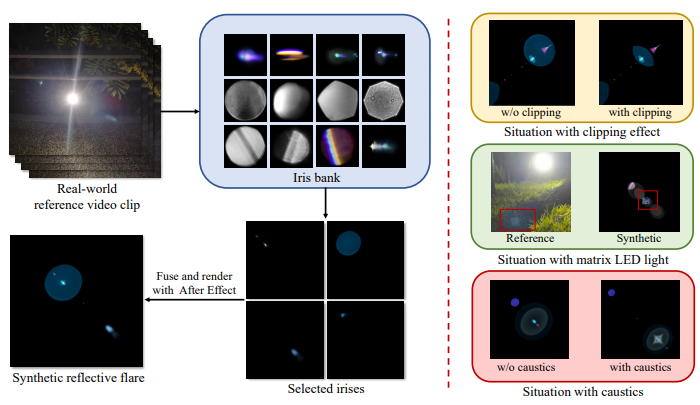}
    \caption{The pipeline of reflective flare synthesis. Since clipping effects and caustics are not obvious in a single image, they captured video clips as references. While synthesizing reflective flares, they first filtered most similar irises from Optical Flares Plug-in’s iris bank. Then, they manually adjusted the position, size, and color of these irises to fit the reference. Finally, these irises are fused to create a reflective flare template in Adobe After Effect. From \cite{flare7kpp}.}
    \label{fig:flare7k_reflective}
\end{figure*}

which, by definition, is the image of a point light source at $(x, y, z)$ formed by a camera with aperture function A. This is the flare image that we desire.

In order to recreate the effect of dust and scratches on a camera lens, they simulated these imperfections by adding random dots and streaks of varying sizes and transparency to a basic model of the lens opening (aperture function), as described in Equation 2. You can see an example of this synthetic lens opening in Figure \ref{fig:wu_scattering} (a), they created a total of 125 different variations of this lens opening.

When we have a specific point of light at a given location (x, y, z) with a single color (wavelength represented by $\lambda$), we can calculate the two aspects of how light is altered as it passes through the lens, which is described in Equation 3 ($\phi_\lambda^S$ and $\phi_\lambda^{DF}$). These calculations are deterministic, meaning we can precisely determine how the point of light will appear in the final image. We refer to this final appearance as the Point Spread Function (PSF) for that particular light source, denoted as $PSF_\lambda$. This PSF is determined by combining the lens opening (A) and the phase shift ($\phi_\lambda$) as specified in Equation 4.

To recreate the effect of a light source spanning the entire visible spectrum, they calculated the $PSF_\lambda$ for all wavelengths ($\lambda$) ranging from 380nm (violet) to 740nm (red), with intervals of 5nm. This results in a 73-value set of data for each pixel of the PSF, considering the different wavelengths. To get the PSF as observed by an RGB camera sensor, This full-spectrum PSF is left-multiplied by a spectral response function (SRF) which describes the sensitivity of color channel c to wavelength $\lambda$, where c $\in$ ${R, G, B}$ and is represented as a 3×73 matrix. This process allows us to derive the PSF as it would be measured by an RGB camera sensor:

\begin{equation}
\begin{bmatrix}
PSF_R(s, t) \\
PSF_G(s, t) \\
PSF_B(s, t) \\
\end{bmatrix}
= SRF
\begin{bmatrix}
PSF_\lambda=380nm(s, t) \\
\vdots \\
PSF_\lambda=740nm(s, t) \\
\end{bmatrix}
\end{equation}
where $(s, t)$ are the image coordinates. This produces a flare image for a light source located at $(x, y, z)$.

They constructed their dataset of scattering flare by randomly sampling the aperture function A, the light source's 3D location $(x, y, z)$, and the spectral response function (SRF). they also applied optical distortion as barrel and pincushion to augment the PSF RGB images. They generated a total of 3,000 such flare images. An example of the simulated aperture function and the resultant flare image can be shown in Figure \ref{fig:wu_scattering} (b). It is worth noting that the exact details for sampling can be found in the Appendix of Wu et al.'s paper \cite{wuetal}.

\textbf{For the reflective flare}, Due to the difficulty of simulation of this type of flare as it requires an accurate characterization of the optics, which is often unavailable, they captured images of reflective flare in a laboratory setting consisting of a bright light source, a programmable rotation stage, and a fixed-aperture smartphone camera with a f = 13mm lens. The setup is clearly shown in Figure \ref{fig:wu_reflective}.
The setup is insulated from ambient light during capture, The camera is also rotated programmatically so that the light source traces (and extends beyond) the diagonal field of view, from $-75\degree$ to $75\degree$. They captured one HDR image every $0.15\degree$, resulting in 1,000 samples. Adjacent captures are
then interpolated by 2x using the frame interpolation algorithm of Context-aware Synthesis for Video Frame Interpolation \cite{contextawareinterpolation}, giving a total of 2,000 reflective flare images. It is worth noting that Wu et al. \cite{wuetal} introduced the very first test dataset consisting of 20 paired flare-corrupted and flare-free images. However, this dataset is extremely homogeneous and only contains images where the light source doesn't exist in the scene.

\begin{table*}[ht]
    \centering
    \caption{A comparison between our Flare7K++ dataset (Flare7K+Flare-R) and Wu et al.’s dataset. The ‘type (s+r)’ indicates the number of different patterns of scattering (s) flare and reflective (r) flare. In the Flare-R dataset, the three smartphones they used possess 9 rear cameras which result in 9 types of reflective flares. They provide glare annotations that also contain the shimmer effect due to the difficulty of separation. From \cite{flare7kpp}.}
    \begin{tabular}{ccccccccc}
        \toprule
        \multirow{2}{*}{Dataset} & \multicolumn{4}{c}{Statistics} & \multicolumn{4}{c}{Annotations} \\
        \cmidrule(lr){2-5} \cmidrule(lr){6-9}
        & Number & Synthetic & Real & Type & Light Source & Reflective Flare & Streak & Glare \\
        \midrule
        Wu et al. & 5,001 & 3,000 & 2,001 & 2+1 & $\times$ & $\times$ & $\times$ & $\times$ \\
        Flare7K & 7,000 & 7,000 & 0 & 25+10 & $\checkmark$ & $\checkmark$ & $\checkmark$ & $\checkmark$ \\
        Flare-R & 962 & 0 & 962 & 962+9 & $\checkmark$ & $\times$ & $\times$ & $\times$ \\
        \bottomrule
    \end{tabular}
    \label{table:datasets_comparison}
\end{table*}

\subsection{Flare7K}
\begin{figure*}[ht]
    \centering
    \includegraphics[width=\linewidth]{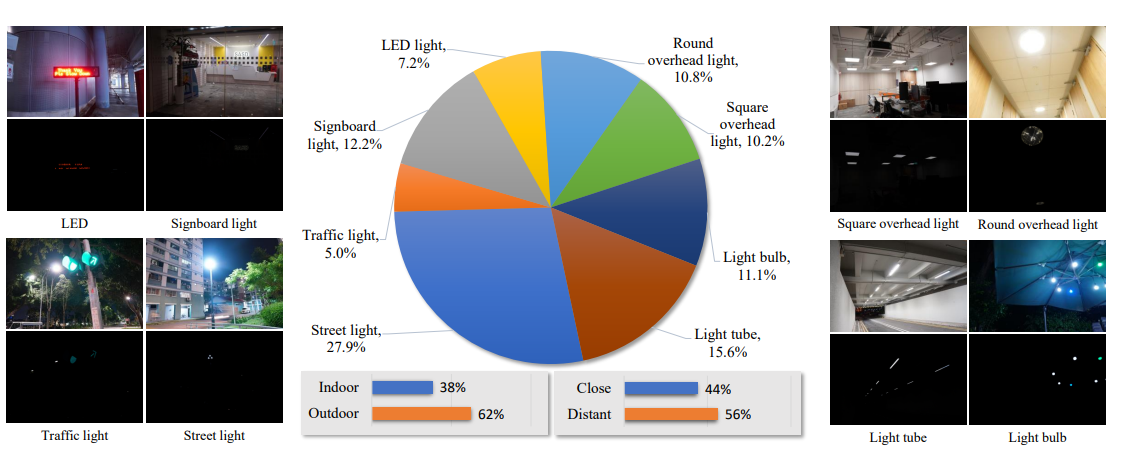}
    \caption{Visualization and distribution of Bracket Flare dataset. It contains different types of light sources in diverse scenes. Based on light source types, they were classified into eight categories with different flare patterns. From \cite{dai2023nighttime}.}
    \label{fig:BracketFlare}
\end{figure*}
So far, the only existing flare dataset is the one proposed by Wu et al. \cite{wuetal}, which is mainly designed for daytime flare removal. Thus, the streak effect and glare effect that commonly exists in nighttime flares are not considered in Wu et al.’s dataset. In terms of nighttime flares. The variety of contamination types on the lens makes it difficult for physics-based methods such as Wu et al.’s approach to collect real nighttime flares by traversing all different pupil functions. This results in the lack of diversity and the domain gap between synthetic flares in Wu et al.’s dataset and the real-world scenes. That's why Dai el al. \cite{flare7k} tried to solve this issue by proposing a new nighttime flare dataset \cite{flare7k} based on hundreds of nighttime flare images with different types of lenses (smartphone and camera) and various light sources as reference images. They also assumed that the flare image can be linearly added to a flare-free image to generate a flare-corrupted image.

They aggregated the captured images and summarized the scattering flares into 25 typical types based on their patterns. Reflective flares, which are associated with the type of the camera's lens group, were also considered. They captured video clips using different cameras as references for simulating reflective flare effects. By referring to these real-world nighttime flare videos, they designed specific flare patterns for each type of camera, resulting in 10 typical types of reflective flares. For each flare type, they generated 200 images with varying parameters like glare radius and streak width, totaling 5,000 scattering flares and 2,000 reflective flares. To utilize how mature the flare rendering has become, they used the Video Copilot Optical Flares plugin in Adobe After Effects. The resulting flare patterns are more diverse and closely resemble real-world nighttime scenarios. A comparison between their dataset \cite{flare7k} and Wu et al.'s \cite{wuetal} dataset in Table 1 reveals that their new dataset offers a wider range of patterns and annotations, which can benefit various applications such as lens flare segmentation and light source extraction.

\textbf{In the scattering flare generation process}, they followed a specific pipeline, as shown in Figure \ref{fig:flare7k}. They broke down the lens flare into four components: shimmer, streak, glare, and the light source. For each of these components, they analyzed parameters such as the glare's radius range and color-distance curve using reference images. They then used Adobe After Effects to create templates for the flare.

For glare synthesis, they examined the relationship between the RGB value of each pixel and its distance from the light source in reference to flare-corrupted images. This relationship was used to create a color correction curve, which was applied to a gradient pattern with the glare's radius to produce the glare effect. To address variations in luminance around the streak, they manually determined the vanishing angle and applied a feathered mask to reduce the glare's opacity in those areas.

When it came to streak synthesis, they departed from a symmetric approach in \cite{Sun_2020_LearnedOpticHDR} as it assumes that the streak is always generated with a 2-point star PSF. However, the streak effect is not usually symmetric, and one side is often much sharper than the other side. They instead manually created masks for each type of streak in Adobe After Effects, allowing for different widths. They used the RGB values of the streak and glare sections to colorize the streak and blur the mask's edges. The blur size was determined based on the curve’s half-life value.

For shimmer synthesis, they utilized a shimmer template in the Optical Flares template and adjusted its parameters to approximate the flare in the image. In areas around the light source where images often experience significant degradation, they employed Adobe After Effect’s default plugin fractal noise and radial blur effects to generate a noise pattern, which was then combined with the shimmer template to create a realistic shimmer effect.

In terms of light source synthesis, they applied thresholding \cite{thresholding} to the flare-corrupted images to identify overexposed tiny shapes. To simulate the glow effect of the light source, they used a plugin called Real Glow in Adobe After Effects on these shapes. The light source region was deliberately made larger than the glare's overexposed part to ensure a realistic visual result. The final step involved adding the light source to the flare using a screen blend mode, preserving the overexposed region's boundaries for a realistic appearance.

\textbf{For reflective flares}, they utilized the same Optical Flares plugin, which included a collection of 51 high-quality iris images. These iris images served as the iris bank. They compared these irises with the reference video and selected the most similar ones \ref{fig:flare7k_reflective}. Then they manually adjusted the size and color of these irises using the Optical Flares plugin. To maintain the proportional relationship between the irises' distances from the light source, they aligned different iris components along a line with proportional distances from the light source. This process resulted in the creation of a reflective flare template.

For specific types of reflective flares, they took into account dynamic triggering mechanisms, including caustics and the clipping effect.

Caustics are optical patterns that occur when light rays refract and converge at a point. In this case, caustics happened when the light source was positioned far from the optical center of the camera lens. To replicate this caustic effect realistically, they utilized a premade template provided by the Optical Flares plugin. This template generated a caustic pattern, a distinctive pattern of light and dark areas, at the center of the iris.

To mimic the dynamic triggering effect associated with caustics, they made the intensity or opacity of the caustics pattern proportional to the distance between the iris and the light source. This meant that as the distance between the iris and the light source changed, the intensity of the caustic effect would also vary accordingly, creating a more authentic representation.

The clipping effect was another phenomenon they considered. It occurred when the path of reflected light was blocked by more than two lenses' apertures. Essentially, it resembled the intersection of two irises. When the distance between the iris and the light source exceeded a certain threshold (the clipping threshold), they initiated a process to remove parts of the iris. To do this, they used another iris as a masking tool. This secondary iris served solely as a mask and was not visible in the final image. This technique allowed them to simulate the effect where portions of the iris were blocked or clipped by the camera's optical components, resulting in a more realistic rendering of the flare.

In nighttime situations, matrix LED light is common which is a distinctive form of flare known as a lattice-shaped reflective flare can manifest in captured images. This specific type of flare occurs due to the arrangement of the LED lights and their impact on the camera lens. To imitate this effect, they synthesized some irises in the shape of the lattice as shown in Figure \ref{fig:flare7k_reflective}.

\textbf{Due to limited nighttime flare removal test dataset}, they created their own synthetic dataset and gathered their own real nighttime flare dataset for evaluation using full-reference image quality assessment metrics.

For synthetic test data collection, they followed the same pipeline discussed earlier. In this approach, neither flare images nor flare-free images were part of their training dataset. To create more realistic synthetic test data, they captured images of the same scenes using both a Huawei P40 smartphone camera (left uncleaned to introduce natural lens flare) and a Sony $\alpha$ 6400 camera with a Sigma 16mm F1.4 lens \cite{sigmalens} (which did not introduce lens flare). Using the smartphone's flare-corrupted images as a reference, they synthesized flare images and merged them with images taken by the professional camera to create pairs of flare-free and flare-corrupted images. They generated a total of 100 such pairs for testing.

For real test data collection, they aimed to replicate the flares found in nighttime scenes, which are typically caused by smudges or scratches on the camera lens (or windshield for nighttime driving). To simulate these real-world flares, they used their fingers and a cloth to introduce common smudges on the camera's front lens. They then used lens tissue to lightly clean the lens, resulting in flare-free images. However, this cleaning process could still cause slight misalignment between the paired images. To address this, they manually aligned the paired images and obtained 100 pairs of real-world flare-corrupted and flare-free images for their test dataset. It's important to note that because the ground truth images may still contain minor flares due to lens defects, the evaluations using these real-world paired images serve as reference points and may not fully reflect the actual performance of flare removal methods.
\subsection{Flare7K++}
Since the synthetic dataset did not include complex degradations caused by diffraction and dispersion in the lens system, a real-world flare dataset, Flare-R \cite{flare7kpp}, was introduced, consisting of 962 real flare patterns. They replicated common lens contaminants encountered in daily use to diversify Flare7K dataset. Various liquids, including water, oil, ethyl alcohol, and carbonated drinks, were applied to the lens surface and wiped with fingers and different types of clothing. After each wipe, a new lens flare image was captured. Automatic white balance on phone cameras was disabled, and different-colored lens flare images were obtained by adjusting the color temperature of the light source.

The Huawei P40, iPhone 13 Pro, and ZTE Axon 20 5G phones, each equipped with three rear lenses of varying focal lengths, were used. For each lens, they altered the distance between the camera and the light source and captured approximately 100 images. In total, they amassed 962 flare images, covering a wide range of common scenarios. Table \ref{table:datasets_comparison} shows how effective the Flare7K \cite{flare7k} and Flare7K++ \cite{flare7kpp} is compared to Wu et al.'s \cite{wuetal} dataset.

In contrast to the synthetic Flare7K dataset, obtaining light source annotations for real-captured flares posed challenges. To address this, they trained a network using Flare7K data and its light source annotations to automatically extract light sources from the real flare images. This network was applied to process all images in the Flare-R dataset to obtain light source annotations. In very few instances where reflective flares were not removed, impacting the accuracy of light source extraction, erroneous bright spots were manually removed to refine the light source annotations.

For the Flare7K test dataset, the ground truth may still be influenced by the slight flares brought by the lens’s defects, global PSNR cannot fully reflect the performance of flare removal methods. To address this problem, they manually labeled masks for all streak and glare regions. Masked PSNR can be used to evaluate the restored results in the regions of different components of flares. they named the masked area of glare and streak regions G-PSNR and S-PSNR respectively. All of those evaluation metrics are discussed in further detail in Section \ref{sec:performance review}.

\subsection{Bracket Flare}

To tackle the challenge of creating paired images for removing reflective flares, Dai et al. \cite{dai2023nighttime} introduced a semi-synthetic dataset called Bracket Flare. Reflective flares typically follow the same pattern as the brightest part of a light source. To generate reflective flare examples, they adjusted the camera's exposure settings to capture the patterns of the light source's brightest region. Their approach involved using a professional camera which is a Sony A7 III \cite{sony_a7} mounted on a tripod. They took a series of five photos of the same stationary scene with a 3 EV step (exposure amount) between each shot. From each group of images, they selected the normally exposed image and the one with the lowest exposure to form a pair.

To enhance the dataset's diversity and realism, they carefully selected scenes with various bright light sources that could produce reflective flares. In the end, they curated a dataset comprising 440 pairs of 4K resolution images for training purposes and an additional 40 images for testing. You can get the distribution of the dataset's contents from Figure \ref{fig:BracketFlare}, which illustrates the dataset's diversity, featuring various common flare patterns. However, it is worth noting that this dataset mainly consists of only reflective flare which is much less common than scattering flare. In addition to this, they relied on optical symmetry prior, which means that bright spots of reflective flares and
light sources are always symmetrical around the lens’s optical center. However, this only holds true for many smartphones and a limited number of professional cameras, so it doesn't generalize well for all types of flare

\section{Performance Review}
\label{sec:performance review}
Performance review in the context of flare removal is essential for evaluating the effectiveness of methods and algorithms designed to mitigate or eliminate lens flare artifacts from images. It serves as a critical step in assessing how well a flare removal technique has performed and whether it meets the intended objectives. It is crucial because it provides valuable insights into the quality and reliability of flare removal results.

\subsection{Performance Metrics}
\subsubsection{\textbf{Peek Signal-to-Noise Ratio}}
\label{sec:psnr}
PSNR \cite{psnr} for short, is a commonly used performance metric in image processing tasks, including flare removal. It quantifies the quality of an image by measuring the similarity between the original, unaltered image and a processed or reconstructed image.

The PSNR is calculated using the Mean Squared Error (MSE) between the original image \(I\) and the processed image \(\hat{I}\).

\begin{equation}
\text{PSNR} = 10 \cdot \log_{10}\left(\frac{{\text{MAX}^2}}{{\text{MSE}}}\right)
\end{equation}

Where:\begin{itemize}
    \item \text{MAX} is the maximum possible pixel value in the image. For typical 8-bit grayscale images, MAX is 255. For color images, it is 255 for each channel (R, G, B). 
    \item MSE \cite{mse} is the Mean Squared Error, which quantifies the average squared difference between corresponding pixels in the original \({I}\) and processed images \(\hat{I}\)
\end{itemize}

\begin{equation}
\text{MSE} = \frac{1}{N} \sum_{i=1}^{N} (I_i - \hat{I_i})^2
\end{equation}

Where:\begin{itemize}
    \item N is the total number of pixels.
    \item  \({I_i}\) is the value of pixel at position \({i}\) in the original image.
    \item \({\hat{I_i}}\) is the value of pixel at position \({i}\) in the processed image.
\end{itemize}

If the PSNR is very high (above 30 dB), it suggests that the processed image is nearly identical to the original, and the quality is excellent. However, it's important to note that while PSNR provides a quantitative measure of image quality, it may not always align perfectly with human perceptual quality. Other metrics and visual inspection are often used in conjunction with PSNR to comprehensively evaluate the success of flare removal methods.

\subsubsection{\textbf{Structural Similarity Index Measure}}
SSIM \cite{ssim} for short, is a performance metric commonly used in image processing tasks, including flare removal. It quantifies the similarity between two images, typically an original image and a processed or reconstructed image. SSIM assesses not only pixel-level differences but also structural and perceptual aspects, making it a valuable metric for evaluating the quality of flare removal. The SSIM index is computed using three main components: luminance, contrast, and structure. 

\textbf{Luminance} ($\mu_x, \mu_y$): Represents the brightness or intensity of the images. SSIM evaluates whether the luminance of the processed image matches that of the original image.

\textbf{Contrast} ($\sigma_x^2, \sigma_y^2$): Measures the variation in pixel intensity within each image. SSIM assesses whether the contrast in the processed image is consistent with that of the original image.

\textbf{Structure} ($\sigma_{xy}$): Captures the structural similarity between the images. It considers how patterns, edges, and textures align between the processed and original images.

\begin{equation}
    \text{SSIM}(x, y) = \frac{{(2\mu_x\mu_y + C_1)(2\sigma_{xy} + C_2)}}{{(\mu_x^2 + \mu_y^2 + C_1)(\sigma_x^2 + \sigma_y^2 + C_2)}}
\end{equation}
Where:\begin{itemize}
    \item $\mu_x$ represents the mean (average) of image x.
    \item $\mu_y$ represents the mean (average) of image y.
    \item $\sigma_x^2$ represents the variance of image x.
    \item $\sigma_y^2$ represents the variance of image y.
    \item $\sigma_{xy}$ represents the covariance between images x and y.
    \item $C_1$ and $C_2$ are the small constants added to stabilize the formula.
\end{itemize}
The SSIM index produces a value between -1 and 1, where a value of 1 indicates that the two images being compared are identical in terms of luminance, contrast, and structure. If it is close to 1 then it suggests a high degree of similarity between the two images. On the other hand, if it is equal to -1 it indicates dissimilarity or a significant difference between the images. A value close to 0 implies that the images are essentially uncorrelated or dissimilar.

By considering not only pixel-level differences but also structural and perceptual similarities, SSIM provides a more comprehensive assessment of image quality in the context of flare removal compared to metrics like PSNR.

\subsubsection{\textbf{Learned Perceptual Image Patch Similarity}}
LPIPS for short, is a performance metric used in image processing tasks, including flare removal. LPIPS assesses the perceptual similarity between two images by considering their structural and textural characteristics. It is particularly useful when evaluating image quality from a human perception perspective. LPIPS is based on deep neural networks that have been trained to measure the perceptual similarity between images. The mathematical representation of LPIPS involves the use of deep neural networks, specifically convolutional neural networks (CNNs). A lower LPIPS score indicates greater similarity between the images, while a higher score suggests greater dissimilarity. The score is obtained by comparing features extracted from different layers of the neural network, capturing information about structural and textural details in the images.
\begin{equation}
\text{LPIPS}(x, y) = \sum_{i=1}^{N} w_i \cdot d_i(x, y)
\end{equation}

Where:\begin{itemize}
    \item $d_i(x, y)$: The feature distance between images x and y at layer $i$ of the deep neural network. Can be MSE, Euclidean distance, cosine distance, or a combination of them
    \item $w_i$: Weights associated with layer $i$ of the network.
    \item $N$: Total number of layers in the network
\end{itemize}

Unlike traditional metrics that only consider pixel-level differences as PSNR, LPIPS evaluates the perceptual similarity between the original and processed images, taking into account structural and textural information. This makes it well-suited for evaluating flare removal techniques that aim to maintain the overall visual quality of the image while reducing flare artifacts.
\begin{table*}[ht]
\caption{Comparison of different methods on the Flare7K++ \cite{flare7kpp} real test dataset.}
\small '*' denotes models with reduced parameters due to the limited GPU memory.
\centering
\begin{tabular}{c|ccc|cc}
\hline
Metric/Method & PSNR & SSIM & LPIPS & G-PSNR & S-PSNR \\ \hline
Input & 22.561 & 0.857 & 0.0777 & 19.556 & 13.105 \\
Wu et al. (U-Net) \cite{wuetal} & 24.613 & 0.871 & 0.0598 & 21.772 & 16.728 \\
Flare7K (U-Net) \cite{flare7k} & 26.978 & 0.890 & 0.0466 & 23.507 & 21.563 \\ 
Flare7K++ (U-Net) \cite{flare7kpp, unet} & 27.189 & \underline{0.894} &  0.0452 & 23.527 & 22.647 \\
Flare7K++ (HINet) \cite{flare7kpp, HINET} & 27.548 & 0.892 & 0.0464 & \textbf{24.081} & \textbf{22.907} \\
Flare7K++ (MPRNet*) \cite{flare7kpp, MPRNet} &  27.036 & 0.893 & 0.0481 & 23.490 & 22.267 \\
Flare7K++ (Restormer*) \cite{flare7kpp, Restormer} & \underline{27.597} & \textbf{0.897} & \underline{0.0447} & 23.828 & 22.452 \\
Flare7K++ (Uformer) \cite{flare7kpp, uformer} & \textbf{27.633} & \underline{0.894} & \textbf{0.0428} & \underline{23.949} & \underline{22.603} \\ \hline 
\end{tabular}
\label{tab:table1}
\end{table*}

\subsubsection{\textbf{Extra Metrics}}
The Glare PSNR and Streak PSNR are specialized performance metrics used in flare removal introduced in Flare7K++ \cite{flare7kpp} dataset, both of which employ the Peak Signal-to-Noise Ratio (PSNR). These metrics focus on evaluating the quality of specific components within an image: glare and streaks, respectively. Glare PSNR calculates the PSNR specifically for the glare portion of the image. It quantifies how well a flare removal method has reduced or eliminated the glare artifacts while preserving the rest of the image's quality. Streak PSNR, on the other hand, computes the PSNR for streak artifacts within an image. It measures how well an algorithm has handled streak-like flare artifacts, such as elongated lines or streaks that may result from intense light sources. 

It's important to note that while Glare PSNR and Streak PSNR are valuable metrics, they have a limitation. These metrics require the availability of glare and streak masks for the test data. In practice, obtaining accurate glare and streak masks may not always be feasible or straightforward and sometimes may not be accurate. This limitation can make it challenging to apply G-PSNR and S-PSNR in all scenarios. Nevertheless, when glare and streak masks are available, these metrics provide valuable insights into the performance of flare removal methods, allowing for targeted evaluation of algorithms.

\subsection{Quantitative Analysis}
Due to limited paired flare-corrupted and flare-free image datasets, the currently available dataset contains a relatively small number of samples. Furthermore, each test dataset has a uniform distribution and can sometimes be limited to only one type of flare. It is worth noting that most of the published papers conduct a user study to compare different methods in addition to quantitative analysis. In this section, we will analyze the state-of-the-art methods on currently used benchmarks\footnote{Due to unavailability of some checkpoints, not all of the variations for each method is considered in all of the conducted quantitative analysis.}.

\subsubsection{\textbf{Wu et al.'s Test Dataset}}
\begin{table}[t]
\centering
\resizebox{\columnwidth}{!}
{
    \begin{tabular}{|c|c|c|c|c|c|c|}
    \hline
    \multirow{2}{*}{Method} & \multicolumn{3}{c|}{Synthetic} & \multicolumn{3}{c|}{Real} \\
    \cline{2-7}
     & PSNR & SSIM & LPIPS & PSNR & SSIM & LPIPS \\ \hline
    Input & 20.449 &  0.832 &  0.1570 & 18.573 & 0.782 & 0.1083 \\
    Wu et al. (U-Net) \cite{wuetal, unet} & \textbf{28.475} &  \textbf{0.9344} & \textbf{0.0502} & \textbf{25.551} & \textbf{0.849} & \textbf{0.0616} \\
    Flare7K++ (Uformer) \cite{flare7kpp, uformer} & \underline{22.911} &  \underline{0.880} & \underline{0.0964} & \underline{19.570} &  \underline{0.794} & \underline{0.1016} \\
    Bracket Flare (MPRNet) \cite{dai2023nighttime, MPRNet} & 20.454 & 0.832 & 0.1567 & 18.579 & 0.782 & 0.1082 \\
    \hline
    \end{tabular}
}
\caption{Comparison for different methods on Wu et al.'s \cite{wuetal} test dataset (Both synthetic and real).}
\label{table:wu et al test dataset}
\end{table}

\begin{table}[t]
\centering
\resizebox{\columnwidth}{!}
{
    \begin{tabular}{|c|c|c|c|c|c|}
    \hline
    Method & PSNR & SSIM & LPIPS & G-PSNR & S-PSNR  \\ \hline
    Input & 22.770 & 0.921 & 0.0601 & 18.804 & 13.927 \\
    Flare7K++ (Uformer) \cite{flare7kpp, uformer}& \textbf{29.498} & \textbf{0.962} & \textbf{0.0210} & \textbf{26.685} & \textbf{24.686} \\
    Bracket Flare (MPRNet) \cite{flare7kpp, MPRNet}& \underline{22.994} & \underline{0.922} & \underline{0.0578} & \underline{18.862} & \underline{14.309} \\
    \hline
    \end{tabular}
}
\caption{Comparison for different methods on Flare7K++ \cite{flare7kpp} synthetic test dataset.}
\label{table:flare7k synthetic test dataset}
\end{table}

Wu et al. \cite{wuetal} introduced two test datasets, one is real containing 20 samples, and the other one is synthetic containing 38 samples. From Table \ref{table:wu et al test dataset}, we can see the Wu et al.'s approach demonstrates state-of-the-art performance on this benchmark. However, the real test dataset contain only images where light source is not present in the image, which means that it only consist of limited types of flare and does not include scattering flare since all of images are captured during daytime not nighttime. Moreover, the synthetic test dataset does not look like the actual flare in real life.

\subsubsection{\textbf{Flare7K++ Test Dataset}}
Dai et al. \cite{flare7kpp} introduced two extra datasets where one of them is real and the other one is synthetic, each containing 100 samples. Table \ref{tab:table1} shows metrics for different flare removal methods on Flare7K++ real test dataset where Uformer \cite{uformer} demonstrates state-of-the-art performance using Dai et al.'s \cite{flare7kpp} approach in Flare7K++ paper. Table \ref{table:flare7k synthetic test dataset} compares the discussed approaches in this survey on Flare7K++ synthetic test dataset. Unfortunately, as there is no checkpoint available for Wu et al.'s work, we are not able to provide their results. It is worth noting that this benchmark contains mainly scattering flare in nighttime with little to none reflective flare, that is why Bracket Flare approach \cite{dai2023nighttime} nearly gives the same results as the plain input since it is only trained on reflective flare.

\subsubsection{\textbf{Bracket Flare Test Dataset}}
\begin{table}[t]
\centering
\resizebox{\columnwidth}{!}
{
    \begin{tabular}{|c|c|c|c|c|}
    \hline
    Method & PSNR & SSIM & LPIPS & Masked PSNR  \\ \hline
    Input & 37.30 &  0.99 & 0.025 & 21.68 \\
    Wu et al. (U-Net) \cite{wuetal, unet} & 26.13 & 0.895 & 0.055 & 21.89 \\
    Flare7K++ (Uformer) \cite{flare7kpp, uformer} & 28.29 & 0.911 &  0.052 & 22.62 \\
    Bracket Flare (MPRNet) \cite{dai2023nighttime, MPRNet}& \textbf{48.41} & \textbf{0.994} & \underline{0.004} & \textbf{32.09} \\
    Bracket Flare (Uformer) \cite{dai2023nighttime, uformer}& 47.47 & \underline{0.991} & \textbf{0.003} & \underline{31.57} \\
    Bracket Flare (HINet) \cite{dai2023nighttime, HINET} & 48.03 & \textbf{0.994} & \textbf{0.003} & 30.88 \\
    Bracket Flare (Restormer) \cite{dai2023nighttime, Restormer} & \underline{48.11} & \textbf{0.994} & \underline{0.004} & 31.07 \\
    \hline
    \end{tabular}
}
\caption{Comparison for different methods on Bracket Flare \cite{dai2023nighttime} test dataset.}
\label{table:bracketflare test dataset}
\end{table}

Dai et al. \cite{dai2023nighttime} introduced another synthetic test dataset consisting of 40 samples in Bracket Flare Paper \cite{dai2023nighttime}. This dataset is based on optical center symmetry prior which states that the reflective flare is in the same position as light source but rotated 180 degrees. However, this only holds true for many smartphones and limited number of cameras. Table \ref{table:bracketflare test dataset} shows metrics for the discussed methods on this dataset. We can clearly see that the input gives even a better performance than Flare7K++, this is due to the fact that this dataset contain only reflective flare and does not even include any scattering ones which Flare7K++ network is trained on.\footnote{Dai et al. \cite{dai2023nighttime} retrained a model with the same method as Wu et al. \cite{wuetal}, that is how we reported Wu et al.'s \cite{wuetal} metrics in Bracket Flare benchmark \cite{dai2023nighttime}.}

% It has the highest number of testing samples among all the available datasets which is 100 samples. However, this benchmark has multiple images for the same scene which may not cover all possible scenarios. In addition to this, the ground truth may still have remaining artifacts and is not 100\% accurate due to alignment issues. Nonetheless, it is still used in quantitative analysis to compare different flare removal methods. Table 2 demonstrates the accuracy of different flare removal methods using the discussed performance review metrics where \textbf{bold} indicates the best score and \underline{underline} indicates the second-best score.

% It is worth noting that there are other benchmarks but they are not commonly used. For instance, Wu et al \cite{wuetal} introduced a test dataset of 20 samples. However, the uniform distribution of this test dataset does not make it reliable. In addition to this, it only contains images where the light source is out of the captured scene, so it does not cover all types of flare and does not even include scattering flare.

\section{Challenges and Opportunities}
\label{sec:challenges and opporutnities}
After highlighting the significant progress that has been made, real-world challenges persist due to the inherent complexities and data-dependent nature of lens flare. For this section, we will introduce some of the promising research directions that we believe will help in further advancing flare removal methods.

\subsection{Video Flare Removal}
Up till now, there is not a single paper addressing flare removal for a video, this unexplored domain has a big room for improvement. Extending lens flare removal techniques from static images to video content introduces a set of unique challenges. Videos are inherently dynamic, with constantly changing scenes and lighting conditions. Temporal coherence, or ensuring that flare removal remains consistent and smooth across frames, becomes a significant concern. Techniques must be able to handle the dynamic nature of videos while avoiding flickering or abrupt changes in visual quality.

Additionally, motion compensation is essential in video flare removal. Objects in videos may move across frames, causing flares to interact with dynamic elements. Methods must account for this motion to accurately remove lens flares without introducing artifacts or distortions. This challenge requires the integration of computer vision and video processing techniques to track and analyze motion patterns within the video.

Scalability is another aspect to consider. Videos come in various resolutions, and flare removal algorithms should be scalable to handle both standard and high-definition videos. This scalability is crucial to ensure that lens flare removal can cater to the needs of various video production scenarios, from consumer videos to professional filmmaking.
\subsection{Real Time Flare Removal}
Achieving real-time lens flare removal is a challenging endeavor, especially for applications like live video streaming, where responsiveness is crucial. To address this challenge, researchers need to develop methods that not only effectively remove lens flare but also do so with minimal computational overhead. This necessitates the creation of low-latency methods that can efficiently process each frame of a video stream or image.

One key concern is striking the right balance between computational efficiency and maintaining high-quality results. While faster methods may yield real-time performance, they could compromise the quality of flare removal, potentially leaving behind artifacts or degrading the overall image or video quality. As such, achieving real-time flare removal requires careful optimization. Researchers must explore innovative strategies to maintain both speed and quality.

\subsection{Dataset Collection}
The development and evaluation of effective flare removal algorithms heavily rely on the availability of high-quality datasets. These datasets should encompass diverse flare patterns, representing various scenarios and conditions in which lens flares occur. However, creating comprehensive and representative datasets is a challenging task in itself and can be sometimes labor-intensive. So far, the mainly used datasets are semi-synthetic. 

The used semi-synthetic pipeline has a large room for improvement. For instance, the number of different scattering flare and reflective flare patterns proposed in Flare7K++ are 25 and 10 respectively. However, those numbers are relatively small and do not cover all the types of flare found in real life due to the different luminance, chromatic, and mechanical properties of each camera. Also, different light source properties and camera settings can contribute to the diversity of flare patterns. 

One notable challenge is the limited availability of comprehensive test datasets that accurately represent the wide array of flare scenarios encountered in practice. This limitation hinders the assessment of different flare removal methods, which is why some of the researchers rely on user study as an evaluation metric.

\subsection{Modular-Based Approaches}
Instead of attempting to tackle the flare removal using an end-to-end approach, researchers can explore a more modular approach, breaking down the task into distinct components, each responsible for a specific aspect. This strategy holds significant promise due to the limited availability of comprehensive paired datasets for training end-to-end models. For instance, Qiao et al. \cite{lightsourceguided} made a significant contribution in this direction by recognizing the critical role of light source conditions in generating lens flares. To address this, they introduced a deep learning-based framework that incorporates light source-aware guidance for single-image flare removal. They employed multiple interconnected modules, including a light source detection module, a flare detection module, a flare removal module, and a flare generation module. These modules work in harmony to address various aspects of the flare removal problem. Notably, their approach was primarily designed to work with unpaired image datasets, since the paired datasets were limited at that time. It's important to highlight that, despite its significance, the paper does not have an official implementation.

Exploring component-based or modular-based strategies for flare removal opens up a promising avenue for researchers to delve into lens flare mitigation, allowing for more targeted and specialized solutions. This approach not only leverages domain-specific knowledge but also offers flexibility in adapting to various real-world scenarios and challenges associated with lens flares.
\subsection{Daytime to Nighttime Translation}
Flare artifacts often become more pronounced and challenging to address in nighttime images due to increased contrast between light sources and dark backgrounds. To enhance the performance of deep learning models for flare removal in low-light conditions, it is advantageous to train these models specifically on nighttime imagery. Many existing methods utilize datasets like Flickr24K used in \cite{flickr24k}, which, while valuable, consist of a substantial number of daytime images. As an innovative approach to address this issue, researchers can explore transforming daytime images into nighttime counterparts using image-to-image translation techniques, such as CycleGAN \cite{cycleGAN}.

Daytime-to-nighttime translation involves adapting existing datasets by simulating nighttime conditions, effectively expanding the available training data. By leveraging image translation methods, researchers can introduce the characteristics and challenges associated with nighttime imagery into their training process.

Exploring daytime-to-nighttime translation techniques in the context of flare removal holds significant potential for advancing the field. Researchers are encouraged to investigate and innovate in this area, as it offers a promising avenue for improving the performance and adaptability of deep learning models in addressing flare artifacts under low-light conditions.

\subsection{Multi-Stage Approach}
One common observation in the output of existing flare removal methods is the presence of residual artifacts that resemble noise, particularly noticeable around streaks in scattering flares. Researchers interested in advancing flare removal techniques can explore the concept of multi-stage methods as a potential solution.

In a multi-stage approach, the image processing pipeline is divided into several sequential stages, with each stage addressing a specific aspect of flare removal. By breaking down the problem into these smaller, focused modules, it becomes possible to tackle artifact reduction more effectively. For instance, one stage can be dedicated to detecting and localizing streak artifacts, while another can focus on their removal, and another one to remove any remaining noise or artifacts. This division of labor allows for greater precision and control over the removal process, potentially leading to significantly improved results.

Encouraging further exploration in flare removal is vital for advancing the field. Researchers are encouraged to investigate and innovate in this area, as it holds the promise of substantially enhancing the quality of flare-free images and addressing persistent artifact challenges.

\section{Conclusion}
In this work, we provide a comprehensive survey of the multifaceted domain of lens flare removal. We delve into the complex optics behind flare formation and categorize the diverse types of flare patterns that can manifest. The numerous factors influencing flare appearance are analyzed in depth, spanning light source attributes, lens features, camera settings, and scene content. The survey extensively reviews the wide range of methods proposed for mitigating lens flare artifacts, covering hardware optimization strategies, traditional image processing techniques, and innovative deep learning approaches. Flare removal datasets are examined, shedding light on how paired images were synthesized for training and evaluation. The performance of various methods is compared quantitatively on popular benchmarks, providing insights into the state-of-the-art capabilities and limitations.

The survey highlights that while significant progress has been made, real-world challenges persist due to the inherent complexities and data-dependent nature of lens flare. Striking an effective balance between efficiency, quality, and generalizability remains an open research problem. Promising future directions are suggested. By comprehensively reviewing the field's evolution, benchmarks, and innovations, this survey offers a timely understanding of the complexities of lens flare and the potential of existing solutions. It provides a strong foundation to motivate and guide future efforts aimed at robustly handling flare artifacts across varied imaging scenarios and applications.

% \appendices

% \section*{Acknowledgment}

\ifCLASSOPTIONcaptionsoff
  \newpage
\fi

\small
\bibliographystyle{IEEEbib}
\bibliography{refs}

% \end{IEEEbiography}



\end{document}